\documentclass[useAMS,usenatbib]{mn2e}

\usepackage{times}  % to get nice fonts for PDF
\usepackage{listings}
\usepackage{graphics}
\usepackage{subfigure}
\usepackage{graphicx}
\usepackage{graphicx,xspace}
\usepackage{amssymb}
\usepackage{amsmath}
\usepackage{hyperref}
\usepackage{aas_macros}
\usepackage{lscape}
\usepackage{rotating}
\usepackage{placeins}
\usepackage{natbib}
\usepackage{color}

\newcommand{\galform}{{\sc{galform}}\xspace}

\title[The evolution of the star forming sequence]
{The evolution of the star forming sequence in hierarchical galaxy formation models}
\author[Peter D. Mitchell]{
\parbox[t]{\textwidth}{
Peter D. Mitchell\thanks{\rm E-mail: peter.mitchell@durham.ac.uk }, 
Cedric G. Lacey,
Shaun Cole,
Carlton M. Baugh
}
\vspace*{6pt} \\
Institute for Computational Cosmology, Department of Physics,
University of Durham, South Road, Durham, DH1 3LE, UK.
\vspace*{-0.5cm}}

\begin{document}
\date{\today}
\pagerange{\pageref{firstpage}--\pageref{lastpage}} \pubyear{2013}
\maketitle
\label{firstpage}

\begin{abstract}
It has been argued that the specific star formation rates of star forming galaxies inferred from observational data decline more rapidly below $z = 2$ than is predicted by hierarchical galaxy formation models. 
We present a detailed analysis of this problem by comparing predictions from the \galform semi-analytic model with an extensive compilation of data on the average star formation rates of star-forming galaxies.
We also use this data to infer the form of the stellar mass assembly histories of star forming galaxies.
Our analysis reveals that the currently available data favour a scenario where the stellar mass assembly histories of star forming galaxies rise at early times and then fall towards the present day.
In contrast, our model predicts stellar mass assembly histories that are almost flat below $z = 2$ for star forming galaxies, such that the predicted star formation rates can be offset with respect to the observational data by factors of up to $2-3$.
This disagreement can be explained by the level of coevolution between stellar and halo mass assembly that exists in contemporary galaxy formation models.
In turn, this arises because the standard implementations of star formation and supernova feedback used in the models result in the efficiencies of these process remaining approximately constant over the lifetime of a given star forming galaxy.
We demonstrate how a modification to the timescale for gas ejected by feedback to be reincorporated into galaxy haloes can help to reconcile the model predictions with the data.
\end{abstract}


\section{Introduction}
\label{Introduction}

Understanding the star formation history of the Universe represents an important goal of contemporary
astronomy, both in theoretical modelling and from observations of the galaxy population. 
Traditionally, the main diagnostic used to characterise the cosmic star 
formation history is the volume averaged star formation rate (SFR) density \cite[e.g.][]{Lilly96,Madau96,Hopkins06}. 
This quantity encompasses the 
combined effect of all the physical processes that are implemented in a given theoretical model of galaxy formation. The lack of a 
complete theory of how these processes operate within galaxies means that these models are 
typically designed to be flexible, utilising simple parametrisations with adjustable 
model parameters. The cosmic star formation rate density, along with other global diagnostics used to 
assess the plausibility of a given model, is sensitive to all of these model parameters. 
Hence, while simply selecting a set of parameters to define a viable model is already challenging,
the problem is compounded by the possibility of degeneracies between different model parameters. 
This has prompted the use of statistical algorithms as tools to explore and identify the allowed
parameter space of contemporary galaxy formation models \citep{Bower10,Henriques13,Lu13a,Mutch13,Ruiz13}.

An alternative to attempting to ``solve'' the entire galaxy formation problem from the top down is to try 
to find observational diagnostics that are sensitive to some specific physical processes but not to 
others. A promising area in this regard revolves around the discovery of a 
correlation between the star formation rate (SFR) and the stellar mass of star forming galaxies, 
forming a sequence of star forming galaxies \cite[e.g.][]{Brinchmann04, Noeske07, Daddi07, Elbaz07}. 
This is most convincingly demonstrated in the Sloan Digital Sky Survey \cite[SDSS; ][]{York00} which 
exhibits a clear star forming sequence with relatively small scatter and a power-law slope which is 
slightly below  unity \cite[e.g.][]{Brinchmann04, Salim07, Peng10, Huang12}.

The discovery of the star forming sequence in the local Universe has motivated a series of 
studies which try to establish whether the sequence is in place at higher redshifts \cite[e.g.][]{Noeske07b}. 
This task is challenging because of the difficulties in reliably measuring the star formation rates of 
galaxies. Beyond the local Universe, star 
formation tracers that do not require the application of uncertain dust corrections are typically available
for only the most actively star forming galaxies. This makes it difficult to prove whether or not there is a
clear bimodality between star forming and passive galaxies in the SFR-stellar mass plane. On the other hand, 
it has been demonstrated that star forming and passive galaxies can be separated on the 
basis of their colours over a wide range of redshifts \cite[e.g.][]{Daddi04,Wuyts07,Williams09,Ilbert10,Whitaker11,Muzzin13}. 
This technique can then be combined with stacking in order to measure the average SFR of star forming
galaxies as a function of both stellar mass and redshift. However, the extent to which these convenient colour 
selection techniques can truly separate galaxies that reside on a tight star forming sequence from the remainder of the population remains uncertain.

The significance of the star forming sequence as a constraint on how galaxies grow in stellar mass
has been discussed in a number of studies \cite[e.g.][]{Noeske07b,Renzini09,Firmani10a,Peng10,Leitner12,Heinis14}. 
The small scatter of the sequence implies that the star 
formation histories of star forming galaxies must, on average, be fairly smooth. This has been taken as evidence
against a dominant contribution to the star formation history of
the Universe from star formation triggered by galaxy mergers \cite[e.g.][]{Feulner05a,Noeske07b,Drory08}.
This viewpoint is supported by studies that demonstrate that the contribution from heavily star 
forming objects that reside above the star forming sequence represents a negligible contribution to the number density and only
a modest contribution to the star formation density of star forming galaxies \cite[e.g.][]{Rodighiero11, Sargent12}.

Various studies have shown that a star forming sequence is naturally predicted both by theoretical galaxy formation
models \cite[e.g.][]{Somerville08,Dutton10,Lagos11a,Stringer12,Ciambur13,Lamastra13,Lu13b} and by hydrodynamical 
simulations of a cosmologically representative volume \cite[e.g][]{Dave08,Kannan14,Torrey14}. These models have reported a slope and scatter that is generally 
fairly consistent with observational estimates. However, there have been a number of reported cases where it appears 
that the evolution in the normalisation of the sequence predicted by galaxy formation models is inconsistent with observational estimates 
\cite[e.g.][]{Daddi07,Dave08,Damen09,Santini09,Dutton10,Lin12,Lamastra13,Genel14,Gonzalez14,Kannan14,Torrey14}. This disagreement is often 
quantified by comparing model predictions with observational estimates of the specific star formation rates of galaxies 
of a given stellar mass as a function of redshift. This comparison can also be made for suites of hydrodynamical zoom
simulations which exchange higher resolution for a loss in statistical information for the predicted galaxy formation population
\cite[][]{Aumer13,Hirschmann13,Hopkins13b,Obreja14}. These studies find that it is possible to roughly reproduce the observed specific star 
formation rate evolution, greatly improving over earlier simulations. However, upon closer inspection, it appears that 
in detail, they may suffer from a similar problem to larger simulations and semi-analytical models with reproducing the observed evolution of the star forming
sequence, as noted by \cite{Aumer13}, \cite{Hopkins13b} and \cite{Obreja14}.

It is important to be aware that below $z \approx 2$, comparisons of specific star formation rates
can yield different constraints on theoretical models depending on whether or not star forming galaxies are separated from 
passive galaxies. In principle, if star forming galaxies are successfully separated, any disagreement in the evolution
of their average specific star formation rates between models and observational data should be independent of ``quenching'' caused 
by environmental processes or AGN feedback. Hence, testing the model using the evolution in the normalisation of the star forming
sequence potentially offers a significant advantage, as compared to more commonly used 
diagnostics such as the cosmic star formation rate density, luminosity functions and stellar mass functions. In particular, 
the reduced number of relevant physical processes makes the problem more tractable and offers a way to improve our 
understanding of galaxy formation without having to resort to exhaustive parameter space searches, where arriving at an intuitive 
interpretation of any results can be challenging. This is particularly pertinent if the simple parametrisations used
in theoretical galaxy formation models for processes such as feedback are not suitable to capture the behaviour
seen in the observed galaxy population.

Here, we use the \galform semi-analytic galaxy formation model along with an extensive literature compilation of observations 
of the star forming sequence to explore the shape of the star formation histories of galaxies within the context of a full hierarchical
galaxy formation model. Our aim is to understand the origin of any discrepancies between the predicted and observed evolution 
in the normalisation of the star forming sequence and to demonstrate potential improvements that could be made in the modelling of
the interplay between star formation, stellar feedback and the reincorporation of ejected gas.

The layout of this paper is as follows. In Section~\ref{GALFORM_Section}, we describe the relevant features of the \galform galaxy formation model used for 
this study. In Section~\ref{Sequence_Section}, we present model predictions for the star forming sequence of galaxies and provide a comparison with
a compilation of observational data extracted from the literature. In Section~\ref{SFH_Section}, we compare the predicted stellar mass assembly histories of
star forming galaxies with the average stellar mass assembly histories inferred by integrating observations of the star forming sequence.
We also explore the connection between stellar and halo mass assembly, highlighting the role of different physical processes 
included in the model. In Section~\ref{Modifications_Section}, we explore modifications that can bring the model into better agreement with the data. We discuss our
results and present our conclusions in Section~\ref{Discussion_Section} and Section~\ref{Summary_Section} respectively. Appendix~\ref{MSI_Appendix} provides a detailed
introduction and exploration of how the stellar mass assembly histories of star forming galaxies can be inferred from observations of the star forming sequence.
Appendix~\ref{Invariance_Section} discusses the impact of changing various parameters in the \galform model. Appendix~\ref{H12_section} presents a short analysis of how well the various
models presented in this paper can reproduce the evolution in the stellar mass function inferred from observations.

\section{The \galform Galaxy Formation Model}
\label{GALFORM_Section}

In this section we describe the \galform semi-analytic galaxy formation model, which we use to simulate the assembly of the
galaxy population within the $\Lambda$CDM model of structure formation. The \galform model belongs to a class of galaxy 
formation models which connect the hierarchical assembly of dark matter haloes to galaxies by coupling merger 
trees generated by cosmological N-body simulations of structure formation to a series of continuity equations which control 
the flow of baryonic mass and metals between hot halo gas, cold disk gas and stellar components. These continuity equations 
are designed to encapsulate the effects of physical processes such as the inflow of gas onto galaxy disks by cooling from 
shock heated hydrostatic haloes. Other processes include quiescent star formation in galaxy disks, chemical enrichment of 
the ISM, the ejection of cold gas and metals by supernovae, the suppression of gas cooling by AGN and photoionization 
feedback, galaxy merging and disk instabilities which in turn can trigger both spheroid formation and bursts of 
star formation. A detailed introduction to the model and the associated underlying physics can be found in \cite{Cole00}, 
\cite{Baugh06} and \cite{Benson10}.

Rather than attempting to solve the equations of hydrodynamics to self consistently predict the full spatial distributions
of stars, gas and dark matter within haloes, the equations within \galform can instead be solved by assuming idealised 
density profiles for the various components of a galaxy-halo system. For example, the hot gas and dark matter density profiles 
are assumed to be spherically symmetric and galaxy disks are assumed to follow an exponential surface density profile. Despite
these simplifications, the lack of a complete theory of star formation and feedback processes means that the continuity
equations can only be formulated and solved using a phenomenological approach.  

Several variants of the \galform model have appeared in the literature which feature different parametrisations of the physics 
of galaxy formation. For this study we adopt a slightly modified version of the model presented in \cite{Lagos12} as our 
fiducial model. The model used in \cite{Lagos12} is descended from that originally presented in \cite{Bower06} 
\cite[see also][]{Lagos11a,Lagos11b}. For the fiducial model used in this study, we make a change from an older gas cooling 
model used in \cite{Lagos12}, which evolves according to discrete halo mass doubling events, to the continuous gas cooling 
model presented in \cite{Benson10b}. In the older cooling model \cite[first presented in ][]{Cole00}, the hot gas profile
is reset and the radius within which hot halo gas is allowed to cool onto a disk is reset to zero when haloes double in mass. For this analysis, we found that
this simplification could lead to artificial suppression of cooling inside haloes hosting massive star forming 
galaxies at low redshift. Changing to a continuous cooling model removes this problem but has the side effect of slightly increasing the amount of gas 
available to form stars in the central galaxies of massive haloes. Therefore, in order to recover approximate agreement with 
the local stellar mass function of galaxies, we lower the threshold required for radio mode AGN feedback to be effective at 
suppressing gas cooling in our fiducial model by changing the model parameter $\alpha_{\mathrm{cool}}$ from $0.58$ 
\cite[as in][]{Lagos12} to $1.0$. All of the models used in this study use merger trees extracted from the Millennium dark 
matter N-body simulation \citep{Springel05}\footnote{Data from the Millennium and Millennium-II simulations are available on a 
relational database accessible from \url{http://galaxy-catalogue.dur.ac.uk:8080/Millennium} .}. A description of the merger 
tree construction can be found in \cite{Jiang14} and \cite{Merson13}.

\subsection{Star formation, supernova feedback and gas reincorporation}
\label{galform_feedback}

We now give a more detailed introduction to the treatment of several physical processes included in \galform that are particularly relevant to this study.
Firstly, our fiducial \galform model uses the empirical star formation law presented in \cite{Blitz06}, which has the form

\begin{equation}
\Sigma_{\mathrm{SFR}} = \nu_{\mathrm{SF}} f_{\mathrm{mol}} \Sigma_{\mathrm{gas}},
\label{blitz_eqn}
\end{equation}

\noindent where $\Sigma_{\mathrm{SFR}}$ is the surface density of star formation rate, $\Sigma_{\mathrm{gas}}$ is the total surface density
of cold gas in the galaxy disk, $f_{\mathrm{mol}}$ is the fraction of cold hydrogen gas contained in the molecular phase 
and $\nu_{\mathrm{SF}}$ is the inverse of a characteristic star formation timescale. $\nu_{\mathrm{SF}}$ is constrained directly 
using observations of local galaxies and is set to $0.5 \, \mathrm{Gyr^{-1}}$ for our fiducial model \citep{Lagos11a}. $f_{\mathrm{mol}}$ 
is calculated using an empirical relationship which depends on the internal hydrostatic pressure of galaxy disks \citep{Blitz06, Lagos11a}.

Secondly, the effects of supernova feedback are modelled by expelling cold gas from galaxy disks over each timestep as stars are formed. The outflow
rate is parametrised as a function of the disk circular velocity at the half mass radius, $V_{\mathrm{disk}}$, and is given by

\begin{equation}
\dot{M}_{\mathrm{ej}} = \psi \, (V_{\mathrm{disk}} / V_{\mathrm{hot}}) ^ {- \alpha_{\mathrm{hot}}},
\label{outflow}
\end{equation}

\noindent where $V_{\mathrm{hot}}$ and $\alpha_{\mathrm{hot}}$ are numerical parameters and $\psi$ is the star formation rate. 
It should be noted that these quantities refer to the outflow and star formation rates integrated over the entire galaxy disk. 
The outflow rate is, by convention, characterised in turn by the dimensionless mass loading factor, 
$\beta_{\mathrm{ml}} \equiv \dot{M}_{\mathrm{ej}} / \psi$.
Unlike the parameters included in the prescription for star formation, $\alpha_{\mathrm{hot}}$ and $V_{\mathrm{hot}}$ are treated
as free numerical parameters and are set in order to reproduce  the observed local galaxy luminosity functions \citep{Bower06}. For our fiducial model,
$V_{\mathrm{hot}}$ is set to $485 \, \mathrm{km \, s^{-1}}$ and $\alpha_{\mathrm{hot}}$ is set to $3.2$, as in the \cite{Bower06} and \cite{Lagos12} models.

All of the gas that is expelled from the galaxy disk is then added to a reservoir of ejected gas which, in turn, is reincorporated at the virial 
temperature back into the hot gas halo at a rate given by

\begin{equation}
\dot{M}_{\mathrm{hot}} = \alpha_{\mathrm{reheat}} M_{\mathrm{res}} / t_{\mathrm{dyn}},
\label{reincorporation}
\end{equation}

\noindent where $\alpha_{\mathrm{reheat}}$ is a numerical parameter, $M_{\mathrm{res}}$ is the mass of gas in the reservoir and 
$t_{\mathrm{dyn}}$ is the dynamical timescale of the halo. For our fiducial model, $\alpha_{\mathrm{reheat}}$ is set to $1.26$.
Once gas is reincorporated back into the halo, it is free to cool back onto the galaxy disk. Hence, gas can be recycled many times 
over the lifetime of a given halo before finally being converted into stars.

\subsection{Quenching processes}

This cycle of gas accretion, cooling, star formation, gas expulsion and reincorporation can be disrupted in \galform through a 
number of different physical processes which we briefly outline here. The focus in this study is on actively star forming 
galaxies which are unaffected by these processes. Quenching mechanisms are therefore not the primary focus of our analysis
as, to first order, they change which galaxies populate the star forming sequence, not the position of the sequence in the star formation rate versus stellar mass plane. 
Nonetheless, it is still important to recognise the conditions under which a given model galaxy will drop out of the samples
of star forming galaxies which form the basis of this study.

Firstly, galaxies that form inside dark matter haloes which are accreted onto larger haloes become satellite galaxies. Satellites
are assumed to lose their hot gas reservoirs to the hot gas halo of the host dark matter halo as a result of ram pressure
stripping. Consequently, once a satellite
uses up its cold disk gas to form stars, it will become permanently quenched. We note that the instantaneous removal of the hot
gas haloes of satellites is, at best, a crude representation of the environmental processes such as ram pressure stripping. 
A more detailed stripping model has been explored in \galform \citep{Font08} but inclusion of this would have only 
a minimal impact on the model central star forming galaxy population which will be the focus of this study.

Secondly, as the mean density of the Universe drops towards the present day, radiative cooling timescales for hot gas inside 
haloes grow longer. In the past, this mechanism was the key for theoretical galaxy formation models to match the observed break
at the bright end of the galaxy luminosity function. However, after improved cosmological constraints favoured a higher 
universal baryon fraction, it was demonstrated that this mechanism could no longer fully explain the break \cite[e.g.][]{Benson03}.
Instead, feedback associated with active galactic nuclei (AGN) is invoked as the primary mechanism responsible for quenching 
massive central galaxies in the current generation of galaxy formation models \cite[e.g.][]{DeLucia04, Bower06,Croton06,Somerville08}. AGN feedback 
in \galform is implemented by assuming that cooling from the hot gas halo is completely suppressed if {\it a)} the halo is
in a quasi-hydrostatic cooling regime and {\it b)} the radiative cooling luminosity of the halo is smaller than the AGN 
luminosity multiplied by an efficiency factor. For more details see \cite{Bower06}.

\section{The Star Forming Sequence Of Galaxies}
\label{Sequence_Section}

In this section we first present the relationship between specific star formation rate and stellar mass predicted by our fiducial \galform model over a range of redshifts.
We then explain how we separate star forming and passive model galaxies at different redshifts. We also present a compilation of observational
data that describe how the average specific star formation rate of star forming galaxies depends on redshift and stellar mass. Finally, we compare our model predictions
with the observational data.

\subsection{The star forming sequence in \galform}
\label{star_forming_sequence_galform}

\begin{figure*}
\begin{center}
\includegraphics[width=40pc]{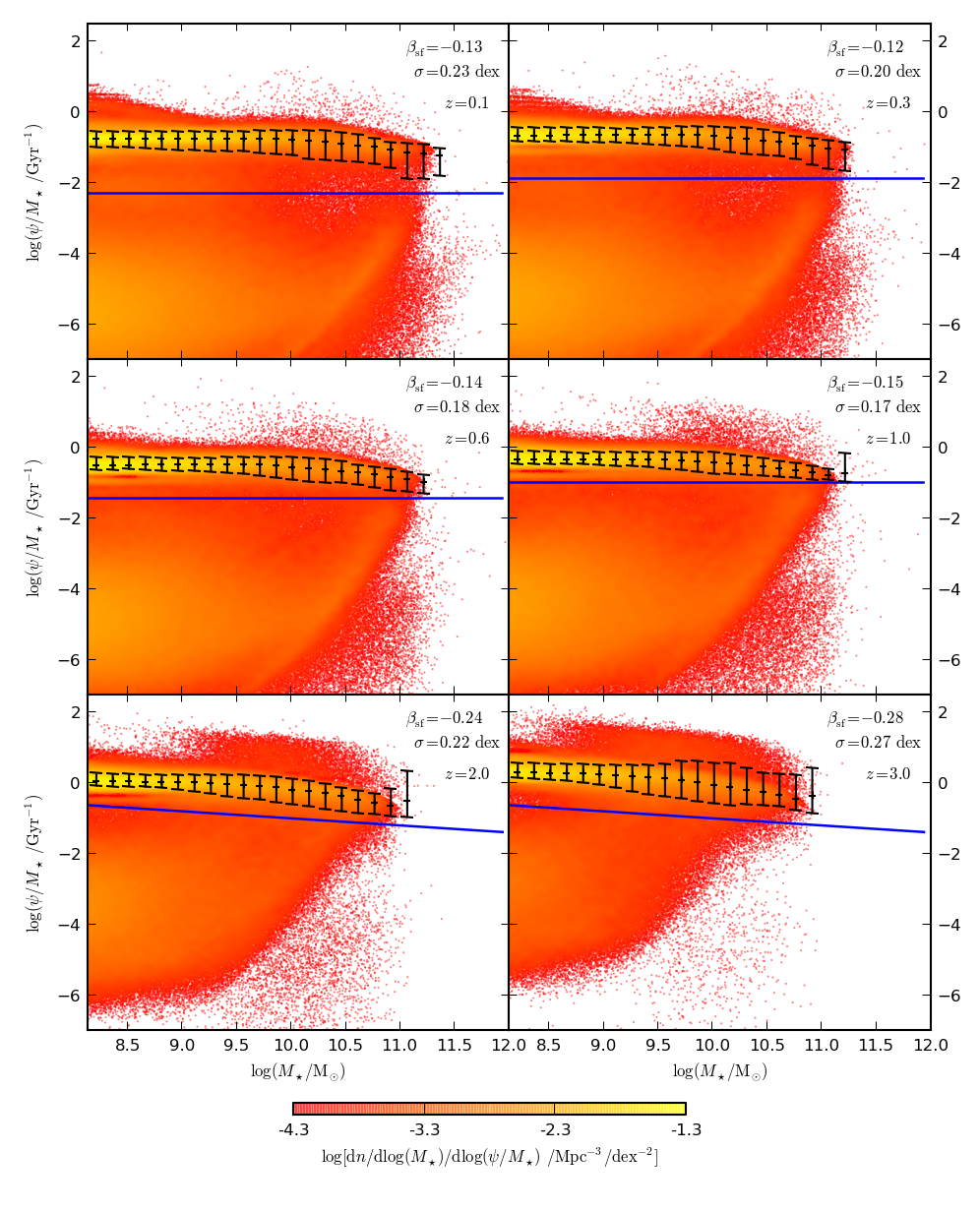}
\caption{Specific star formation rate plotted as a function of stellar mass for all galaxies from our fiducial \galform model. 
Each panel corresponds to a different redshift as labelled.
The coloured points represent individual model galaxies and the point colours are scaled logarithmically with the local number density of galaxies in each panel, from red at low density to yellow at high density.
The corresponding number densities are indicated by the colour bar at the bottom of the figure.
The blue lines show our cut between star forming and passive galaxies for each redshift.
The black points and corresponding error bars show the median, $10^{\mathrm{th}}$ and $90^{\mathrm{th}}$ percentiles of the distribution in the specific star formation rates of star forming galaxies, binned as a function of stellar mass.
$\beta_{\mathrm{sf}}$ is the slope of a power-law fit to the medians of the distribution for star forming galaxies.
$\sigma$ quantifies the average scatter and is defined as half of the mean central $68\%$ range of the distribution for star forming galaxies. }
\label{ssfr_m_evo_Lagos12}
\end{center}
\end{figure*}

Fig.~\ref{ssfr_m_evo_Lagos12} shows the distribution of specific star formation rate against stellar mass in our 
fiducial \galform model for a selection of redshifts. We choose to show individual galaxies as points, coloured by 
the logarithmic density of points at a given position in the plane. For reference, the number of galaxies shown in 
each redshift panel is of order $10^6$.
The most obvious feature that can be seen in Fig.~\ref{ssfr_m_evo_Lagos12} is a strong sequence of star forming galaxies 
that extends over several decades in stellar mass. Outliers that reside above this sequence do exist but are rare, 
becoming slightly more prevalent towards higher redshifts. Passive galaxies reside below the sequence, with a broad 
distribution of specific star formation rates at a given stellar mass. 

For the remainder of this study, we choose to focus on the star forming galaxies that reside either on or above 
the star forming sequence. We separate passive galaxies by applying a power-law cut that evolves with redshift. The 
division is shown as solid blue lines in Fig.~\ref{ssfr_m_evo_Lagos12}. The exact position and slope of the power-law 
cuts are fixed by hand in order to best separate the star forming sequence from the locus of passive galaxies 
that can be seen stretching diagonally across the plane for the most massive passive galaxies at lower redshifts. 
Although this is a subjective process, we find that our results are, in general, insensitive to the precise 
location of the cut because of the strong bimodality in the distribution.
This is demonstrated by the fact that the $10^{\mathrm{th}}$ percentiles of the distribution of star forming 
galaxies do not reside close to our dividing line between star forming and passive galaxies in most cases. The exception to 
this is seen at $z=1$ where the locus of massive passive galaxies joins onto the star forming sequence, making
it difficult to objectively separate star forming from passive galaxies at $M_\star \approx 10^{11} \mathrm{M_\odot}$.

To characterise the slope and normalisation of the star forming sequence seen in Fig.~\ref{ssfr_m_evo_Lagos12},
we adopt the convention from \cite{Karim11} who use a power-law fit of the following functional form,

\begin{equation}
\psi / M_\star = c \, \left(\frac{M_\star}{10^{11} \mathrm{M_\odot}}\right)^{\beta_{\mathrm{sf}}},
\label{karim_fit}
\end{equation}

\noindent where $\beta_{\mathrm{sf}}$ is the slope of the sequence and $c$ sets the normalisation. We also define the 
scatter in the star forming sequence, $\sigma$, as half of the mean value over stellar mass bins of the central $68\%$ range in the 
distribution of $\log_{10}(\psi / M_\star \mathrm{ / {Gyr}^{-1} })$, calculated for each bin in stellar 
mass. The scatter, $\sigma$, and best fitting power-law slope, $\beta_{\mathrm{sf}}$, to the star forming sequence are 
labelled for each panel shown in Fig.~\ref{ssfr_m_evo_Lagos12}. 
We find that the slope steepens from $\beta_{\mathrm{sf}} \approx -0.13$ at $z \leq 1$ up to $\beta_{\mathrm{sf}} = -0.28$ at $z=3$. 
This range of slopes lies comfortably within the range of slopes that are reported by observational studies 
(see Appendix~\ref{MSI_Obs_Section}). The mean scatter, $\sigma$, does not vary strongly 
with redshift below $z=3$ and is typically $\approx 0.2 \, \mathrm{dex}$. The increase to $0.27 \,\mathrm{dex}$ 
at $z=3$ can be attributed to an increased abundance of outlying galaxies that reside above the star 
forming sequence at this redshift. Finally, we note that the normalisation of the star forming sequence 
can be seen to increase by roughly an order of magnitude over $0 < z < 3$. We explore this in greater 
depth in Section~\ref{ssfr_model_obs_comp_section}.

Compared to the results reported for the model from \cite{Dutton10}, the star forming sequence predicted by 
our fiducial model has a larger intrinsic scatter by $\Delta \sigma \approx 0.1 \, \mathrm{dex}$. 
\cite{Dutton10} explain that they expect their model to under-predict the scatter because their model features 
a simplified treatment of the mass assembly histories of dark matter haloes, neglecting various aspects of the 
hierarchical galaxy formation process that are included in \galform. On the other hand, we note that the 
hydrodynamical simulations presented in \cite{Torrey14} and \cite{Obreja14} predict a larger scatter of 
$\approx 0.3 \, \mathrm{dex}$. This could reflect a failing of the simplified treatment of physical processes used in
\galform when compared to a full hydrodynamical simulation. 
The larger scatter reported by \cite{Torrey14} is consistent with the upper limit on the intrinsic scatter 
typically reported from observational studies \cite[e.g][]{Noeske07b, Whitaker12} but it is difficult to 
accurately assess the true uncertainty on the star formation tracers used in these studies. For the 
purposes of this study, the scatter in our model is small enough to be consistent with the observational upper limit 
and from here on, we focus instead on the slope and normalisation of the star forming sequence.

Compared to our fiducial model, the slope of the star forming sequence decreases more slowly with redshift 
in the \cite{Dutton10} model, varying from $\beta_{\mathrm{sf}} = -0.04$ at $z=0$ to $-0.1$ at $z=3$. This slope 
is slightly shallower than predicted by our fiducial model. This could potentially be explained by the lack 
of any quenching or starburst processes in the \cite{Dutton10} model. The slope of $-0.2 < \beta_{\mathrm{sf}} < -0.1$ 
predicted by the model presented in \cite{Lamastra13} is consistent with our fiducial model although this 
somewhat unsurprising given the many similarities between the two models. On the other hand, the 
hydrodynamical simulations presented in \cite{Torrey14} report a slope of $-0.05 < \beta_{\mathrm{sf}} < 0.0$ which is 
more similar to \cite{Dutton10}. \cite{Obreja14} also find a slope consistent with $\beta_{\mathrm{sf}} = 0$.

\subsection{The star forming sequence inferred from observations}

\begin{table*}
\begin{center}
\begin{tabular}{|ccccc|}
\hline
Source & Redshift& Selection& SF cut& Tracer \\
\hline

\cite{Noeske07}&      0.2-1.1& K&            blue colour/$24 \mathrm{\mu m}$ detection&    $24 \mathrm{\mu m}$+UV/Em Lines \\
SDSS DR7       &      0.08&    r&            sSFR-$M_\star$ distribution&         $H_\alpha$ \\
\cite{Pannella09}&     1.5-2.5& BzK&          sBzK&                        Radio \\
\cite{Oliver10}&      0-2&     Optical&      template fitting&            $70/160 \mathrm{\mu m}$ \\
\cite{Magdis10}&      3&       LBG&          blue colour&                 UV (corrected) \\
\cite{Peng10}&        0-1&     Optical&      blue colour&                 SED fitting \\
\cite{Rodighiero10}&  0-2.5&    $4.5 \mathrm{\mu m}$& blue colour/$24 \mathrm{\mu m}$ detection& FIR \\ 
\cite{Karim11}&       0.2-3&   $3.6 \mathrm{\mu m}$& blue colour&         Radio \\
\cite{Huang12}&       0&       HI / r&       HI detection/blue colour&    SED fitting \\
\cite{Lin12}&         1.8-2.2&       BzK&          sBzK&                        UV (corrected) \\
\cite{Reddy12}&       1.4-3.7& LBG&          blue colour&                 $24 \mathrm{\mu m}$+UV \\
\cite{Whitaker12}&    0-2.5&   K&            (U-V/V-J) cut&                 $24 \mathrm{\mu m}$+UV \\
\cite{Bauer13}&       0.05-0.32& r&          $H_\alpha$ flux/$EW$&        $H_\alpha$ \\
\cite{Stark13}&       4-7&     LBG&          blue colour&                 UV (corrected) \\
\cite{Wang13}&        0.2-2&   K&            SFR-$M_\star$ distribution&          SED fitting / FIR \\
\cite{Gonzalez14}&    4-6&     LBG&          blue colour&                 SED fitting \\

\end{tabular}
\caption{List of the sources of the observed average specific star formation rates of star forming galaxies, $\langle \psi / M_\star \rangle (M_\star,z)$, which we extract from the literature.
We list the source, redshift range or median redshift, galaxy selection technique, the subsequent star forming galaxy selection technique, and the tracer used to estimate the instantaneous star formation rate.
For LBG-selected samples, it should be noted that the initial galaxy selection technique is strongly biased towards blue star forming galaxies, so typically no additional cut to separate star forming galaxies is performed.
SDSS DR7 data are taken from the public webpage \protect\url{http://www.mpa-garching.mpg.de/SDSS/DR7/}, which corresponds to an update of the \protect\cite{Brinchmann04} analysis.
For \protect\cite{Karim11}, we use both the star forming galaxy sample presented in their Table 3 as well as the ``active population'' which is shown in their Figure 13 (which uses a bluer colour cut).
The code and observational data used for this compilation are available at \protect\url{http://www.astro.dur.ac.uk/\~d72fqv/average\_sSFR\_SFGs/}.
}
\label{ssfr_mz_comp}
\end{center}
\end{table*}

For this study, we have compiled a set of observational data on the star forming sequence 
for the purposes of providing a comparison with model predictions. Specifically, we have 
compiled the average specific star formation rate of star forming galaxies for bins of 
stellar mass and redshift. Both mean and median star formation rates have been used to 
quantify the average in the literature and we include both quantities in the compilation. 
The list of sources used in the compilation is presented in Table~\ref{ssfr_mz_comp}. We 
include information on the redshift range covered, the initial selection 
technique, the technique to separate star forming galaxies and the star formation rate 
tracer used. We only include observational data sets that have either made an attempt to 
separate star forming galaxies from passive galaxies or have a selection function which 
intrinsically selects only actively star forming objects. Where necessary, we convert 
stellar masses quoted that assume a Salpeter IMF by 
$\Delta \log(M_\star / \mathrm{M_\odot}) = -0.24 \, \mathrm{dex}$ in order to be consistent 
with a Chabrier IMF \citep{Ilbert10,Mitchell13}, which in turn is very similar to the Kennicutt
IMF that is assumed in our model. We do not attempt to correct specific 
star formation rates for IMF variations as we expect both the stellar mass and star formation
rate corrections to approximately cancel in most cases.

It is very important to be aware that the average star formation rate, particularly for 
large stellar masses at low redshift, will depend strongly on the method used to separate 
star forming from passive galaxies. In general, it is not possible in practice to simply make 
the separation based on identifying the star forming sequence in the star formation rate 
versus stellar mass plane. This is only really possible in the local Universe with surveys 
such as the SDSS. Instead, star forming galaxies are often separated using colour selection 
criteria \cite[e.g.][]{Daddi04, Ilbert10}. These issues are particularly pertinent for 
studies that employ stacking techniques, where it is impossible to ascertain whether a 
star forming sequence is really present in the data \citep{Oliver10, Rodighiero10, Elbaz11, Karim11}.

\subsection{Comparing the star forming sequence from \galform with observational data}
\label{ssfr_model_obs_comp_section}

\begin{figure*}
\begin{center}
\includegraphics[width=40pc]{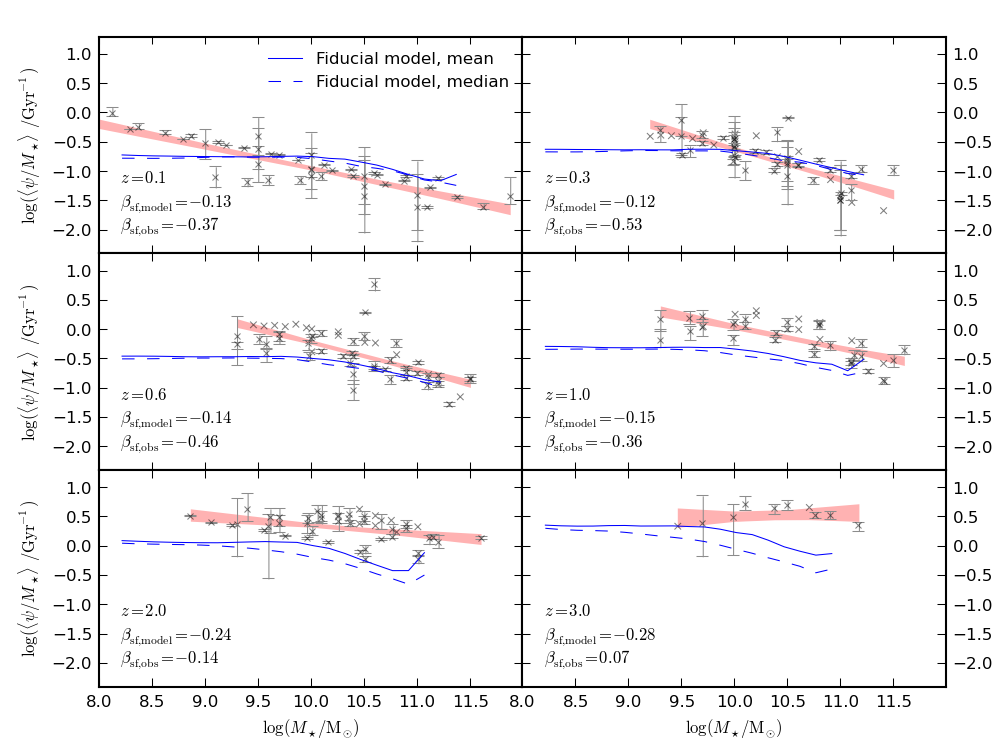}
\caption{The average specific star formation rate of star forming galaxies plotted as a function of stellar mass. 
Each panel corresponds to a different redshift as labelled.
Blue solid and dashed lines show predictions from our fiducial \galform model for the mean and median specific star formation rates respectively.
Grey points show observational estimates of either the mean or median average specific star formation rate of star forming galaxies.
A list of the sources of these observational data points is presented in \protect Table~\ref{ssfr_mz_comp}.
When shown, the corresponding error bars show a lower limit on the statistical uncertainty on the average for each data point.
The shaded region shows the $1 \sigma$ range of power-law fits to the observational data, using a fixed error on the data points of $0.20 \, \mathrm{dex}$.
$\beta_{\mathrm{sf,model}}$ is the best fitting power-law slope to the medians of the distribution predicted by our fiducial model.
$\beta_{\mathrm{sf,obs}}$ is the best fitting power-law slope to the observational data presented in each panel. }
\label{ssfr_m_evo_Lagos_obs_comp}
\end{center}
\end{figure*}

Fig.~\ref{ssfr_m_evo_Lagos_obs_comp} shows the average specific star formation rates of star 
forming galaxies as a function of stellar mass for a selection of redshifts. Observational 
data from the compilation presented in Table~\ref{ssfr_mz_comp} are shown in grey and can be 
compared to the mean and median relations predicted by the our fiducial model. It should be 
noted that the error bars on the observational data points show only a lower limit on the 
statistical uncertainty on each data point. These error bars are only shown for the studies 
where an estimate of this lower limit could be obtained. The error bars do not represent the 
dispersion in the underlying distribution. We attempt to estimate a more realistic 
uncertainty on the average specific star formation rate by measuring the average central $68 \%$ 
range for measurements in all stellar mass bins containing more than two data points. We find that the 
uncertainty on the specific star formation rate estimated in this way is 
$0.20 \, \mathrm{dex}$. The pink shaded regions shown in 
Fig.~\ref{ssfr_m_evo_Lagos_obs_comp} then enclose the set of best fitting power laws to the data 
within a $1 \sigma$ range, assuming $0.2 \, \mathrm{dex}$ errors in each mass bin.

Given the large systematic uncertainties that are thought to affect stellar mass and SFR 
estimates and that each data set uses a different method to select star forming galaxies, it 
is reassuring that the observational data are fairly self consistent in normalisation within 
each respective redshift panel. There are some outlying data sets however. In general, the 
observations seem consistent with a star 
forming sequence where the average specific star formation rate is modestly anti-correlated 
with stellar mass. We note however that there are significant variations in the slope seen 
between different redshift panels. The best fitting power-law slopes at each redshift vary from 
$\beta_{\mathrm{sf}} \approx -0.4$ at $z=0$ to $\beta_{\mathrm{sf}} \approx 0.1$ at $z=3$. Whether or not this variation 
is driven by an intrinsic shift in the slope of a star forming sequence of galaxies is 
extremely unclear. We explore this issue in more detail in Appendix~\ref{MSI_Obs_Section}.

Comparison of the observed slope with predictions from our fiducial model indicates that the model has a 
slope which is too shallow at low redshift and too steep at high redshift. In addition, compared
to the data, the slope variation with redshift acts in the opposite direction in the model, such that the high 
redshift slope is steeper than the local relation. However, it is difficult to be confident 
whether this truly reflects a flaw in the model or can be explained as a 
result of selection effects. To answer this question satisfactorily would require a 
self-consistent comparison between the model and the data in terms of selection. However, 
this task is made challenging because of the difficulty in predicting accurate colour 
distributions for galaxies from hierarchical galaxy formation models. Historically, various 
\galform models have struggled to reproduce the observed colour distributions of galaxies, 
making it difficult to reproduce observational colour cuts in detail \cite[e.g.][]{Guo13}.

\begin{figure*}
\begin{center}
\includegraphics[width=40pc]{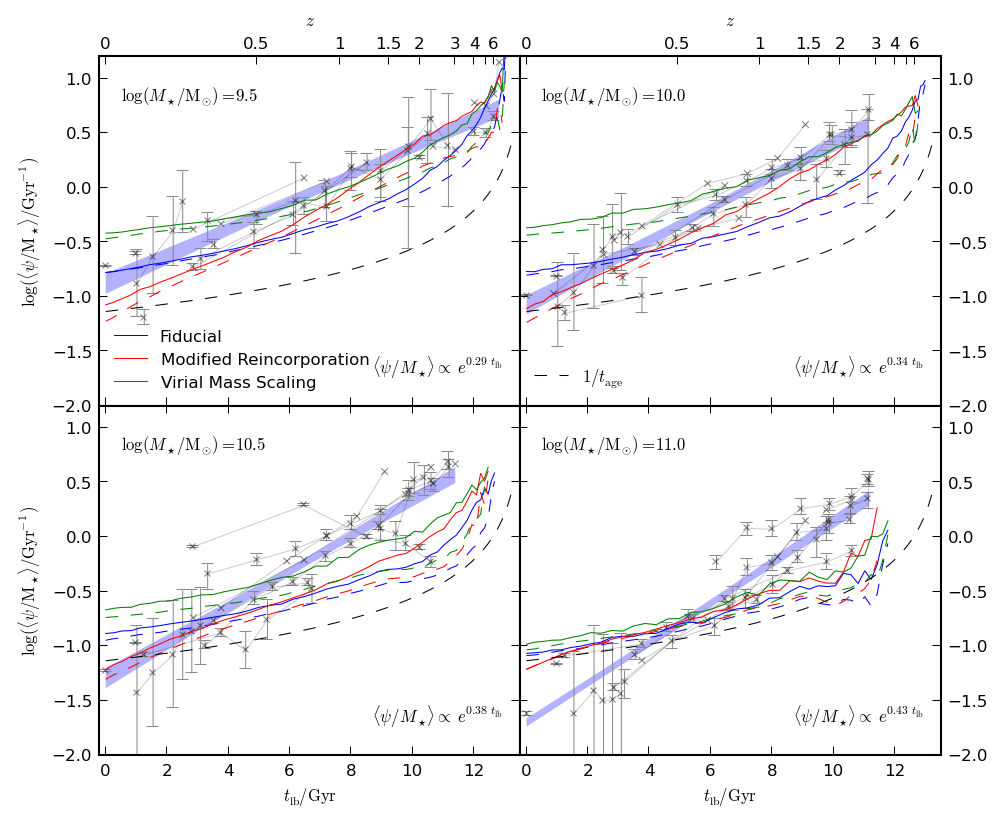}
\caption{The average specific star formation rate of star forming galaxies plotted as a function of lookback time. 
Each panel corresponds to a different stellar mass bin as labelled.
Blue solid and dashed lines show predictions from our fiducial \galform model for the mean and median specific star formation rates respectively.
Red lines show the same information but for the modified reincorporation model.
Green lines show the same information but for a model using the virial mass scaling for the reincorporation timescale proposed by \protect \cite{Henriques13}.
Dashed black lines show the inverse of the age of the universe as a function of lookback time.
Grey points show observational estimates of either the mean or median average specific star formation rate of star forming galaxies.
A list of the sources of these observational data points is presented in \protect Table~\ref{ssfr_mz_comp}.
When shown, the corresponding error bars show a lower limit on the statistical uncertainty on the average for each data point.
Grey points taken from a single observational study are connected by grey lines.
The blue shaded region shows the $1 \sigma$ range of exponential fits to the observational data, assuming a fixed error on the data points of $0.20 \, \mathrm{dex}$.
This shaded region is consistent with, but not identical to, the pink shaded region shown in \protect Fig.~\ref{ssfr_m_evo_Lagos_obs_comp}.
The best fit to the evolution in the observational data is given in each panel.}
\label{ssfr_zm_Lagos_obs_comp}
\end{center}
\end{figure*}

Although it is difficult to draw strong conclusions from comparing the slopes of the 
observed and predicted distributions, it is apparent that the normalisation of the star 
forming sequence evolves more rapidly in the observational data than in the model. This problem is best 
viewed by plotting the evolution of the average specific star formation rate of star forming 
galaxies as a function of lookback time for selected stellar mass bins, which we show in 
Fig.~\ref{ssfr_zm_Lagos_obs_comp}. In general, we find that the observational data are 
consistent with exponential evolution in $\langle \psi / M_\star \rangle$ with lookback time. The 
best fit to the observational data in each panel gives 
$\langle \psi / M_\star \rangle \propto e^{a \, t_{\mathrm{lb}}}$ where $a$ is found to 
vary between $0.29$ and $0.43$. The variation in $a$ is such that the average specific star 
formation rate drops more rapidly with time in the highest mass bins shown. Although there 
is some scatter at a given redshift, the data within each mass bin appear to be mostly self 
consistent in normalisation at a given redshift. Repeating the same process as for 
Fig.~\ref{ssfr_m_evo_Lagos_obs_comp}, we estimate the uncertainty on the average specific 
star formation rates and again find that the uncertainty is approximately $0.20 \, \mathrm{dex}$.

In contrast to the trend that emerges from the observational compilation, our fiducial 
model (blue lines) predicts slower evolution than the data (until higher redshifts, where the evolution 
in the model becomes steeper than an extrapolation of the trend seen in the data). This 
behaviour has been seen for various published models in the literature 
\cite[e.g.][]{Dave08,Damen09,Dutton10,Gonzalez14,Torrey14} and the origin of the discrepancy is the subject 
of the remainder of this paper. Finally, at this stage we note that the evolution in the fiducial \galform 
model scales very closely with the inverse of the age of the universe at a given time, 
$t_{\mathrm{age}}$. We return to this point in Section~\ref{Explain_SFH_Section}.

\section{The Stellar Mass Assembly Of Star Forming Galaxies}
\label{SFH_Section}

It is clear from Fig.~\ref{ssfr_zm_Lagos_obs_comp} that the specific star formation
rates of galaxies at a fixed stellar mass evolve more slowly with redshift in our fiducial
\galform model than is implied by the observational data. However, it should be noted that the galaxy population 
which is probed at each redshift for a given stellar mass bin will not be the same; star forming 
galaxies grow in stellar mass before becoming quenched and consequently dropping out of the star 
forming samples which we consider. This complicates the interpretation of 
Fig.~\ref{ssfr_zm_Lagos_obs_comp} with regard to understanding the physical origin
of any flaws in the model. 

It is therefore worthwhile to search for a another way to characterise the evolution
of star forming galaxies which traces only a single population across cosmic time. 
One way to achieve this is to try to infer the stellar mass assembly histories 
of star forming galaxies by tracing how they grow in stellar mass as they evolve along 
a star forming sequence. This technique has already appeared in various guises in the 
literature \cite[e.g.][]{Drory08,Renzini09,Leitner11,Leitner12,Heinis14}. 
From here on in, we adopt the terminology of \cite{Leitner12} and refer to this technique 
as Main Sequence Integration (MSI).

In this section, we explore the origin of the discrepancy between the predicted and observed
evolution in the specific star formation rates of star forming galaxies shown in Fig.~\ref{ssfr_zm_Lagos_obs_comp}
We start by making use of the MSI technique to compare the predicted and observationally inferred stellar
mass assembly histories of galaxies that are still star forming at $z=0$. We then go on to discuss why
the predicted stellar mass assembly histories have a particular form, connecting the halo assembly
process with the way stellar feedback is implemented in our model.

An introduction to the MSI technique is presented in Appendix~\ref{MSI_Introduction}. An exploration of how well MSI can recover
the predicted stellar mass assembly histories calculated in our model can be found in Appendix~\ref{MSI_galform}. Details of the observational compilation of 
measurements of the star forming sequence which we use for this study can be found in Appendix~\ref{MSI_Obs_Section}.

\subsection{Comparing the inferred stellar mass assembly histories of star forming galaxies with model predictions}
\label{sfh_model_obs_comp}

\begin{figure*}
\begin{center}
\includegraphics[width=40pc]{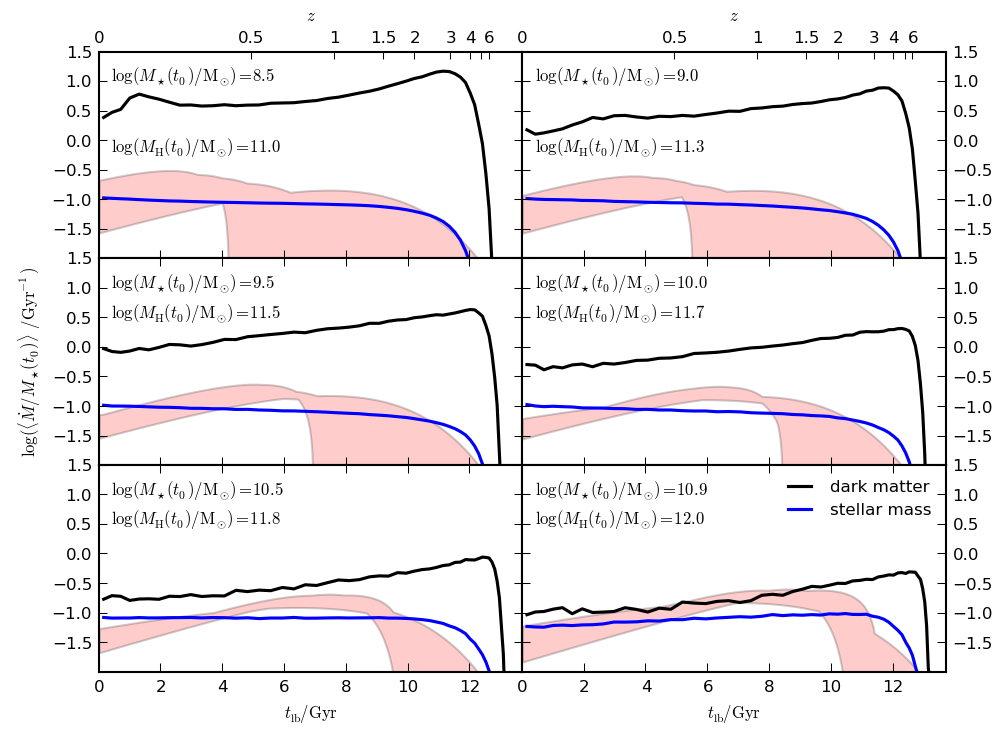}
\caption{The mean mass assembly histories of galaxies that are central and star forming at $z=0$, plotted as a function of lookback time.
Blue lines show predictions for the mean stellar mass assembly histories for the main stellar progenitors of central galaxies, taken directly from our fiducial \galform model.
Black lines show the corresponding dark matter halo mass assembly histories of the progenitor haloes that host the main stellar progenitors of central galaxies at $z=0$. 
These curves are rescaled by $f_{\mathrm{b}} \equiv \Omega_{\mathrm{b}}/\Omega_{\mathrm{M}}$ to show the baryonic accretion rate onto these haloes.
Model galaxies are binned by their stellar mass at $z=0$, with each panel corresponding to a different mass bin.
The median $z=0$ stellar mass in each bin is labelled in each panel.
The corresponding median $z=0$ dark matter halo mass in each stellar mass bin is also labelled.
The filled pink region shows the range of stellar mass assembly histories that are inferred by applying the MSI technique to observational data from the literature.}
\label{hfh_sfh_comp}
\end{center}
\end{figure*}

In Fig.~\ref{hfh_sfh_comp}, we compare the predicted average stellar mass assembly histories of central galaxies 
that are still star forming at $z=0$ from our fiducial \galform model with the results of applying MSI to the observational compilation 
presented in Appendix~\ref{MSI_Obs_Section}. To facilitate a comparison 
with the average stellar mass assembly histories obtained using the MSI technique for a given starting 
mass, model galaxies are binned by their stellar mass at $z=0$.
By default in this study, stellar mass assembly histories are obtained by tracing back the main 
stellar progenitor of each $z=0$ central star forming galaxy. We define the main stellar progenitor 
as the most massive stellar progenitor traced between each consecutive pair of output times. The 
impact of this choice (as compared to summing over all possible progenitors) is discussed in Appendix~\ref{MSI_galform}.
We have chosen to plot 
$\langle \dot{M}_\star / M_\star(t_0) \rangle$ in order to eliminate dispersion associated with the 
finite width of the stellar mass bins which we use ($\Delta \log(M_\star \, / \mathrm{M_\odot}) = 0.5 \, \mathrm{dex}$). 
The remaining dispersion therefore reflects the intrinsic scatter in our model in the shape of stellar 
mass assembly histories of galaxies that are central and star forming at $z=0$.
The choice to include only galaxies that are central at $z=0$ is made in order to minimise the
impact of any environmental effects. When comparing to mass assembly histories inferred from observations (which include a 
combination of satellite and central galaxies), the exclusion of satellite galaxies is justified 
by observational results that indicate that the form of the star forming sequence is independent of 
environment \cite[e.g.][]{Peng10}.

Quantitatively, the model predicts stellar mass assembly rates that are broadly 
consistent to within factors of $2$ compared to the data. However, despite the weak constraints provided by MSI in some cases, there 
is a clear qualitative disagreement between the model and the data regarding the shape of the stellar mass 
assembly histories predicted by \galform. In the model, the rate of star formation rises rapidly at early 
times before slowing down to a gradual rise or to a constant level of activity at later times. The observational data instead
favours a scenario where star formation activity builds towards a peak at an intermediate time before dropping
significantly towards $z=0$. This disagreement is consistent with the trend seen in 
Fig.~\ref{ssfr_zm_Lagos_obs_comp} where the specific star formation rates of galaxies in the model are too 
low compared to the data at intermediate times before rapidly rising towards high redshift.

At this stage, it should be noted that our fiducial model is only one specific realisation of \galform with
regard to the various model parameters that can be changed. While these parameters are constrained by requiring that
the model matches local global diagnostics of the galaxy population, it is important to understand whether the disagreement 
between predictions and data seen in Fig.~\ref{ssfr_zm_Lagos_obs_comp} and Fig.~\ref{hfh_sfh_comp} is specific to the
combination of parameters used in our fiducial model. An analysis of this issue
if presented in Appendix~\ref{Invariance_Section}. To summarise, we find that for a given stellar mass at $z=0$, the shapes of the 
average stellar mass assembly histories of central star forming galaxies in \galform are almost entirely invariant 
when changing model parameters relating to star formation, feedback and gas reincorporation. As a consequence,
the disagreements seen in Fig.~\ref{ssfr_zm_Lagos_obs_comp} and Fig.~\ref{hfh_sfh_comp} do indeed seem to
be generic for any model that uses the same parametrisations to represent these physical processes. 
This also helps to explain why similar models find a similar disagreement in other studies.

As well as stellar mass assembly histories, we also show in Fig.~\ref{hfh_sfh_comp} the corresponding
average dark matter halo mass assembly histories of central galaxies from our fiducial \galform model that are 
star forming at $z=0$. We choose to define the dark matter halo mass assembly rate, $\dot{M}_{\mathrm{H}}$, by tracing 
backwards the host halo of the main stellar progenitor (see Appendix~\ref{MSI_galform}). This definition is useful for making 
comparisons with the stellar mass assembly process. However, it should be noted that in some cases this 
definition could deviate from the standard definition of halo mass assembly histories 
where instead the main halo progenitor is traced backwards \cite[e.g][]{Fakhouri10}. To quantify the 
average halo mass assembly rate, we take the mean of the distribution at each lookback time. This choice is made because the 
individual halo assembly histories are very stochastic with respect to our temporal resolution (which is
determined by the number of available outputs from the Millennium simulation). As a 
consequence, we find that only the mean halo mass assembly history integrates to the correct average halo 
mass at $z=0$ while the median does not. Incidentally, this stochasticity is why even the average halo mass assembly histories shown in
Fig.~\ref{hfh_sfh_comp} get visibly noisy towards late times due to a drop in the average rate of
significant accretion events.

\begin{figure*}
\begin{center}
\includegraphics[width=40pc]{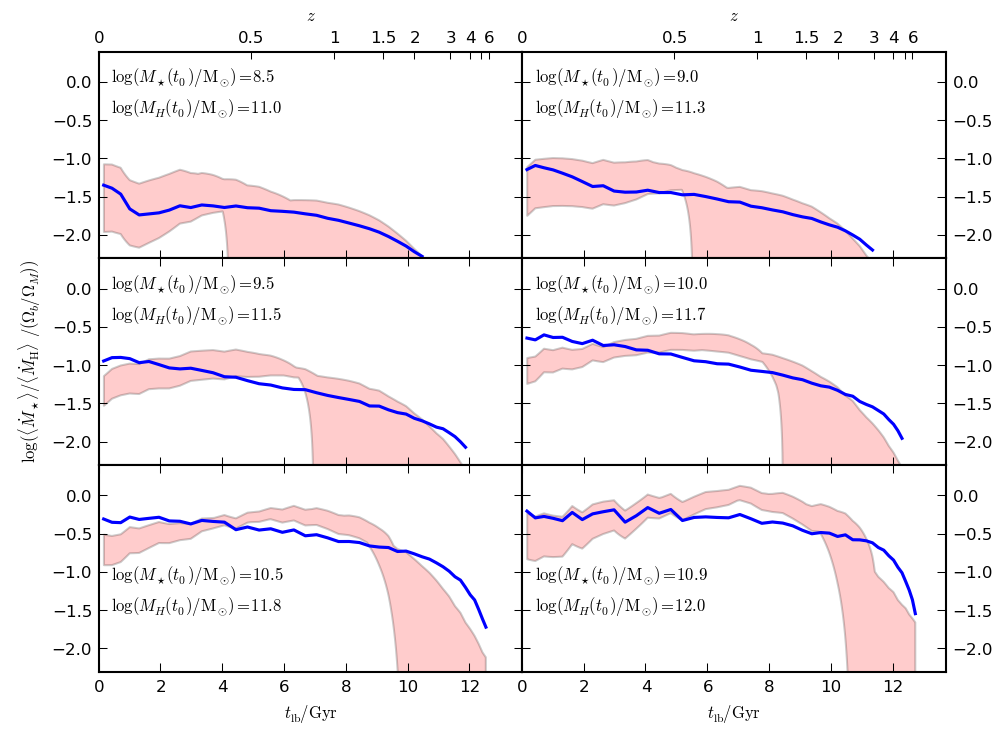}
\caption{The ratio of mean stellar mass assembly rate to mean baryonic halo mass assembly rate for galaxies that are star forming at $z=0$, plotted as a function of lookback time.
Blue lines show predictions from our fiducial \galform model for this ratio for galaxies that are central at $z=0$.
Model galaxies are binned according to their $z=0$ stellar mass with each panel corresponding to a different mass bin.
The median $z=0$ stellar mass in each bin is labelled in each panel.
The corresponding median $z=0$ dark matter halo mass of each stellar mass bin is also labelled.
The filled pink regions show the range in the ratio of galaxy to halo mass assembly rates inferred from observational data.
This is obtained using a combination of stellar mass assembly histories inferred from observational data using the MSI technique and average halo mass assembly histories from \galform.
This assumes that the true ratio between stellar mass and halo mass at $z=0$ is the same as in our fiducial \galform model.}
\label{sfe}
\end{center}
\end{figure*}

From Fig.~\ref{hfh_sfh_comp}, we can begin to understand why there is a disagreement in the stellar mass 
assembly process between our fiducial \galform model and the trends implied by the data. In the model, 
despite the enormous variation in the efficiency of star formation relative to gas accretion between haloes 
of different mass at $z=0$, stellar mass assembly broadly tracks the halo assembly process. Stars start to form as 
soon as their host haloes accrete an appreciable fraction of their final mass and this continues all the way to 
the present day. Differences between the stellar and halo assembly histories do exist however. For example, the stellar 
mass assembly histories do not show the peak at $t_{\mathrm{lb}} \approx 12 \, \mathrm{Gyr}$ which is 
fairly prominent for the haloes. Also, the halo accretion rates fall slowly towards late times after this 
peak, whereas most star forming galaxies form stellar mass at either a constant or slightly increasing rate 
over their lifetimes. However, the decline in the halo mass accretion rates is generally not as steep as the 
rate of decline in star formation rates inferred from the observational data.
This can be seen more clearly in Fig.~\ref{sfe} where we show the ratio of the rates of mean stellar mass 
assembly to halo mass assembly.

In order to broadly reproduce the observed $z=0$ stellar mass or luminosity functions, it is necessary 
that a given galaxy population model, on average, places galaxies of a given stellar mass inside haloes 
of the mass corresponding approximately to the correct abundance. Given that our fiducial \galform model roughly reproduces the local stellar
mass function, the halo assembly histories shown in  Fig.~\ref{hfh_sfh_comp} should therefore correspond 
roughly to the true halo formation histories of real galaxies, for the case of a $\Lambda$CDM universe. 
Adopting this as a working assumption, we also show in Fig.~\ref{sfe} the efficiency of star formation
inferred from observations using MSI if we use the relationship between stellar mass and halo mass at $z=0$ for star forming 
galaxies in our model. This extra step allows us to infer how efficiently
haloes that host star forming galaxies at $z=0$ convert accreted baryons into stars. It can be seen that
in the model, the efficiency of star formation relative to halo gas accretion rises monotonically from 
early times to the present day, whereas the data, in general, favours a scenario where this efficiency 
peaks at some intermediate time for the higher stellar mass bins.

In Appendix~\ref{MSI_Obs_Section}, we discuss how, depending on the slope of the observed star forming sequence, $\beta_{\mathrm{sf}}$, 
the inferred stellar mass assembly of galaxies that are still star forming at $z=0$ can show a downsizing trend, such that
the lower mass star forming galaxies start forming stars later with respect to massive star forming galaxies.
Fig.~\ref{hfh_sfh_comp} shows that any possible downsizing trend suggested by the data is, at best, only weakly 
reproduced by the model. The different stellar mass assembly histories that we infer from the observational data agree 
best with our model for the $\beta_{\mathrm{sf}} = 0.0$ or $-0.1$ cases shown in Fig.~\ref{sfh_obs}. The assembly histories
derived from these bins show the weakest downsizing trend (no downsizing for $\beta_{\mathrm{sf}} = 0.0$) and form a greater 
fraction of stars at early times. It should be noted that the better agreement with the model for these curves 
is not surprising given that the slope of the star forming sequence in the fiducial \galform model is 
$\beta_{\mathrm{sf}} \approx -0.15$. For the opposite extreme case in the data where $\beta_{\mathrm{sf}} \approx -0.5$, the model 
predictions are in dramatic disagreement with the trends implied by the data for low mass galaxies.

Another consequence of a strong downsizing trend is that the stellar 
mass assembly process is significantly delayed relative to the halo assembly process for low mass systems.
Fig.~\ref{hfh_sfh_comp} shows that the 
shape of the mean halo mass assembly histories is only a very weak function of the final halo mass. Therefore, 
any possible downsizing trend that exists purely in the star forming population would have to be caused by a 
physical process which is separate from the growth of the hosting dark matter haloes. For the case where
$\beta_{\mathrm{sf}} \approx -0.4$, such a process would result in the existence of a population of dark 
haloes that have not formed any appreciable amount of stars at intermediate redshifts of $1 < z < 2$. We note 
that the star formation histories presented in \cite{Leitner12}, derived by applying MSI to data from \cite{Karim11} with 
$\beta_{\mathrm{sf}} = -0.35$ and from \cite{Oliver10}, would also have this 
consequence. Reproducing this behaviour in models or simulations would require much stronger feedback (or
the inclusion of another physical mechanism with the same effect) at early times than is typically assumed for galaxies 
that reside within the progenitors of haloes of mass $11 < \log(M_{\mathrm{H}}(t_0) / \mathrm{M_\odot}) < 12$.

\subsection{Explaining the form of stellar mass assembly histories in \galform}
\label{Explain_SFH_Section}

In Section~\ref{sfh_model_obs_comp}, we show that the stellar mass assembly process in our fiducial 
\galform model broadly traces the halo mass assembly process. The closeness in this predicted co-evolution 
appears to be in qualitative disagreement with trends inferred from the star formation rates of galaxies 
inferred from observational data. This leads to the slower evolution in the predicted average specific star 
formation rates of star forming galaxies compared to the observational data seen in 
Fig.~\ref{ssfr_zm_Lagos_obs_comp}. Fig.~\ref{ssfr_zm_Lagos_obs_comp} also demonstrates that this evolution 
in the model closely traces the inverse of the age of the universe, $t_{\mathrm{age}}$, 
such that $\psi / M_\star \propto 1/t_{\mathrm{age}}$. We now consider why the model behaves in this way.

\begin{figure}
\includegraphics[width=20pc]{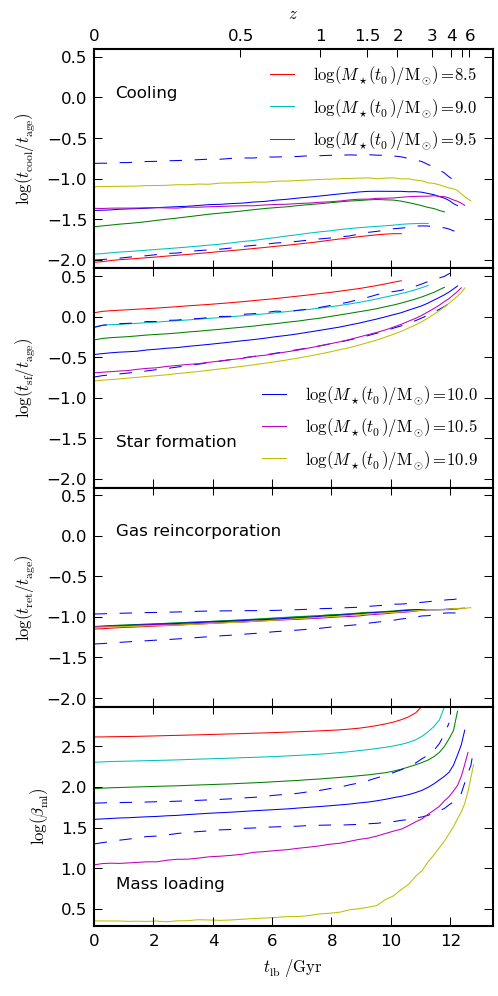}
\caption{The ratio of average characteristic timescales of model galaxies which are central and star forming at $z=0$ to the age of the universe, $t_{\mathrm{age}}$, plotted as a function of lookback time.
Model galaxies are binned according to their $z=0$ stellar mass with each solid coloured line showing the median of the distribution for a different bin.
The blue dashed lines show the $10^{\mathrm{th}}$ and $90^{\mathrm{th}}$ percentiles for the $\log(M_\star(t_0)/\mathrm{M_\odot}) = 10$ bin.
The top three panels each correspond to a different timescale. The bottom panel instead shows the efficiency with which gas is ejected from galaxy disks.
{\it Top:} The characteristic gas cooling timescale, $t_{\mathrm{cool}}$, for hot halo gas to cool onto a galaxy disk.
{\it Second:} The characteristic star formation timescale, $t_{\mathrm{sf}}$, for disk gas to be converted to stars in the absence of feedback.
{\it Third:} The characteristic gas reincorporation timescale,  $t_{\mathrm{ret}}$, for gas ejected by feedback to be reincorporated back into the hot gas halo.
{\it Bottom:} The mass loading factor of outflows, $\beta_{\mathrm{ml}}$.}
\label{timescales}
\end{figure}

{\it i) Cooling timescale:} For the star forming galaxy population which we consider in this study, we expect
the radiative cooling timescales for shock heated halo gas to cool onto galaxy disks to 
be short compared to the age of the Universe at a given epoch. In the top panel
of Fig.~\ref{timescales}, we see that this is indeed the case if we trace backwards the
main stellar progenitors of galaxies that are central and star forming at $z=0$, following
the methodology introduced in Section~\ref{sfh_model_obs_comp}. We define the 
characteristic cooling timescale, $t_{\mathrm{cool}}$, as the time for gas with the mean 
density within the virial radius to cool. Given that this timescale is short, the only three
remaining physical processes in the model which are relevant for the star forming galaxy 
population considered here are star formation, outflows triggered by SNe feedback and the 
subsequent reincorporation of ejected gas back into the hot halo gas component.

{\it ii) Star formation timescale:} The efficiency of star formation can be characterised by the timescale required to consume 
cold disk gas in the absence of feedback. This is given by $t_{\mathrm{sf}} \equiv M_{\mathrm{cold}}(t) /\psi(t)$.
We show the average evolution in this quantity for the star forming population in the
second panel of Fig.~\ref{timescales}, relative to the age of the universe at a given epoch.
It can be seen that this timescale is typically comparable to the age of the universe, 
although there is an order of magnitude variation depending on the time and final stellar 
mass of the galaxies being considered. The consequence of the balance between the star formation
timescale and the age of the Universe is that cooling gas can be effectively converted into stars in a quasi-steady 
state. In practice, the true gas depletion timescale will be significantly shorter than 
$t_{\mathrm{sf}}$ in the model when the mass loading factor of outflows, $\beta_{\mathrm{ml}}$, rises above unity, 
which is typically the case for the galaxies considered here.

{\it iii) Mass loading factor:} In our model, the efficiency of SNe feedback in ejecting cold gas from galaxies 
is characterised by the mass loading factor, $\beta_{\mathrm{ml}}$. The average evolution in $\beta_{\mathrm{ml}}$ 
is shown in the bottom panel of Fig.~\ref{timescales}. As described in 
Section~\ref{galform_feedback}, $\beta_{\mathrm{ml}}$ in our fiducial \galform 
model scales $\propto V_{\mathrm{disk}}^{-3.2}$, where $V_{\mathrm{disk}}$ is the circular 
velocity of the galaxy disk at the half mass radius. The evolution in this quantity over the
lifetime of the star forming galaxies which we consider here is very modest. In general, $V_{\mathrm{disk}}$ rises 
at early times before becoming almost constant at intermediate to late times. This lack of 
evolution in $V_{\mathrm{disk}}$ is primarily driven by the corresponding lack of evolution in the 
circular velocity of the host haloes at the virial radius, $V_{\mathrm{vir}}$.
The strong scaling of $\beta_{\mathrm{ml}}$ with $V_{\mathrm{disk}}$ means that there is a stronger evolution in the
efficiency of feedback with lookback time, particularly for massive galaxies at early times where 
$\beta_{\mathrm{ml}}$ grows significantly above unity. The increased feedback efficiency at early times
explains why the galaxy stellar mass assembly histories do not share the peak at $t_{\mathrm{lb}} \approx 12 \, \mathrm{Gyr}$ seen for
the halo mass assembly histories shown in Fig.~\ref{hfh_sfh_comp}.
At late times, $\beta_{\mathrm{ml}}$ becomes approximately
constant in time for all galaxies, which will result in a fixed modulation of the efficiency in converting 
accreted gas into stars.

{\it iv) Reincorporation timescale:} It is also important to consider how efficiently gas that is ejected by feedback is able to return back 
into the hot gas halo. As described in Section~\ref{galform_feedback}, ejected gas is placed into 
a reservoir of mass $M_{\mathrm{res}}$. This gas then returns to the halo on a characteristic timescale 
given by $t_{\mathrm{ret}}(t) \equiv M_{\mathrm{res}} / \dot{M}_{\mathrm{ret}}$ \citep{Bower06}. In \galform, this 
quantity scales $\propto t_{\mathrm{dyn}}^{-1}$ where $t_{\mathrm{dyn}}$ is the halo dynamical time (see Eqn.~\ref{reincorporation}). 
We characterise the efficiency of gas reincorporation by $t_{\mathrm{ret}} / t_{\mathrm{age}}$. 
The average evolution in this reincorporation timescale for model galaxies that are central and star forming at 
$z=0$ is shown in the third panel of Fig.~\ref{timescales}. This shows that the timescale for reincorporation is 
close to an order of magnitude shorter than the age of the universe at all times. The 
timescale is also almost completely independent of the final stellar or halo mass. This is because the 
halo dynamical time, to first order, depends only on the current mean density of the universe. As the mean density 
of the universe falls with time, so does the timescale for reincorporation.

By combining the picture that is presented in Fig.~\ref{timescales} with 
simple arguments, we now proceed to demonstrate analytically the origin of the behaviour seen in 
Fig.~\ref{ssfr_zm_Lagos_obs_comp} for the predicted evolution of the specific star formation rates of star
forming galaxies. Firstly, we can relate the mean density of a 
halo, $\bar{\rho}_{\mathrm{H}}$ to the circular velocity at the virial radius, $V_{\mathrm{vir}}$, and 
the virial radius, $R_{\mathrm{vir}}$, through

\begin{equation}
\bar{\rho}_{\mathrm{H}} = \frac{3}{4 \pi} \frac{M_{\mathrm{H}}}{R_{\mathrm{vir}}^3} = \frac{3}{4 \pi G^3} \frac{V_{\mathrm{vir}}^6}{M_{\mathrm{H}}^2 } .
\label{density_halo}
\end{equation}

\noindent This can be rearranged into

\begin{equation}
M_{\mathrm{H}} \propto {\bar{\rho}_{\mathrm{H}}}^{-0.5} V_{\mathrm{vir}}^3.
\label{mhalo}
\end{equation}

\noindent The average density of a halo can be related to the mean density of the universe, $\bar{\rho}$, by

\begin{equation}
\bar{\rho}_{\mathrm{H}} =  \Delta_{\mathrm{v}} \bar{\rho},
\label{collapse}
\end{equation}

\noindent where $\Delta_{\mathrm{v}}$ is the mean overdensity given by the spherical collapse model 
\citep{Gunn72}. For the simplified case of an $\Omega_{\mathrm{M}} = 1$ universe, 
$\Delta_{\mathrm{v}} = 18 \pi^2$ and $\bar{\rho} \propto 1/{t_{\mathrm{age}}^2}$. In this case, we can write

\begin{equation}
M_{\mathrm{H}} \propto t_{\mathrm{age}} \, V_{\mathrm{vir}}^3.
\label{mhalo_scaling}
\end{equation}

\noindent As discussed earlier, evolution in $V_{\mathrm{vir}}$ over the lifetime of a given galaxy is weak, particularly at intermediate to 
late times. We can therefore make the approximation that $V_{\mathrm{vir}}$ is constant with time, yielding

\begin{equation}
\dot{M}_{\mathrm{H}} \propto V_{\mathrm{vir}}^3.
\label{dbdt_mhalo_scaling}
\end{equation}

If we temporarily ignore gas reincorporation and assume that the star formation, freefall and cooling 
timescales of a halo are all short, then balancing the rate of accretion of gas to star formation and gas
ejection gives

\begin{equation}
f_{\mathrm{b}} \dot{M}_{\mathrm{H}} = \dot{M}_\star + \dot{M}_{\mathrm{ej}},
\label{ode}
\end{equation}

\noindent where $\dot{M}_{\mathrm{ej}}$ is the rate of ejection of gas mass by feedback and $f_{\mathrm{b}}$
is the baryon fraction relative to dark matter. 
Eqn.~\ref{ode} can be rewritten in terms of the dimensionless mass loading factor, $\beta_{\mathrm{ml}} = \dot{M}_{\mathrm{ej}} / \psi$, yielding

\begin{equation}
f_{\mathrm{b}} \dot{M}_{\mathrm{H}} = \dot{M}_\star \, (1+\beta_{\mathrm{ml}}/(1-R)),
\label{ode_sfr}
\end{equation}

\noindent where $R$ is the fraction of gas recycled back into the ISM as a result of stellar evolution, which is 
assumed to be a constant so that $\dot{M}_\star = (1-R) \psi$. At this stage we note that supernova feedback in our fiducial \galform model is parametrised as 
$\beta_{\mathrm{ml}} \propto V_{\mathrm{disk}}^{-3.2}$ (see eqn.~\ref{outflow}). Therefore, in the regime under consideration where $V_{\mathrm{disk}}$ does not 
evolve with time, $\beta_{\mathrm{ml}}$ is constant. In this regime, Eqn.~\ref{ode_sfr} can be integrated to give

\begin{equation}
f_{\mathrm{b}} M_{\mathrm{H}} = M_\star \, (1+\beta_{\mathrm{ml}}/(1-R)).
\label{ode_int}
\end{equation}

\noindent If we then substitute the scalings from eqns \ref{mhalo_scaling} and \ref{dbdt_mhalo_scaling} into \ref{ode_int} and \ref{ode_sfr}
and then divide \ref{ode_sfr} by \ref{ode_int}, we find that the specific stellar mass assembly rate is given by

\begin{equation}
\frac{\dot{M}_\star}{M_\star} = \frac{\dot{M}_{\mathrm{H}}}{M_{\mathrm{H}}}  = \frac{1}{t_{\mathrm{age}}}.
\label{ssfr_scaling_basic}
\end{equation}

\noindent We note that \cite{Stringer12} obtain the same result, where they argue that this behaviour is generic in the 
regime where the halo mass assembly process is approximately self-similar.

In Fig.~\ref{ssfr_zm_Lagos_obs_comp}, it can be seen that the evolution of the average specific star formation rates 
in our fiducial \galform model closely tracks the inverse of the age of the universe at a given epoch. By following 
the derivation of Eqn.~\ref{ssfr_scaling_basic}, it can be seen that this behaviour will naturally emerge if
$V_{\mathrm{vir}}$ and $\beta_{\mathrm{ml}}$ remain approximately constant with lookback time. We note that although
this is true for the majority of the lifetimes of star forming galaxies in our fiducial model, Fig.~\ref{timescales} 
shows that the situation changes for $\beta_{\mathrm{ml}}$ at early times. This explains why the efficiency of converting 
accreted gas into stars, as shown in Fig.~\ref{sfe}, rises rapidly at early times.

The derivation of Eqn.~\ref{ssfr_scaling_basic} ignores the reincorporation of gas after 
ejection by feedback. This is actually a very poor approximation given that Fig.~\ref{timescales} shows that $\beta_{\mathrm{ml}}$ is 
always $\gg 1$ over the lifetime of the galaxies which we consider. In addition, the bottom panel of Fig.~\ref{timescales} 
shows that the reincorporation timescale is typically between a factor of $6$ and $20$ shorter than the age of the universe.
Combined, these two features of the model mean that gas reincorporation will be highly significant in shaping the predicted star
formation histories of star forming galaxies. Therefore, it is clear that Eqn.~\ref{ode} needs to be 
modified in order to account for the fact that the gas will typically have been recycled between the galaxy disk and the 
halo many times before forming into stars. To incorporate this effect, Eqn.~\ref{ode} can be rewritten as

\begin{equation}
f_{\mathrm{b}} \dot{M}_{\mathrm{H}} + \dot{M}_{\mathrm{ret}} = \dot{M}_\star + \dot{M}_{\mathrm{ej}} = \dot{M}_\star \, (1+\beta_{\mathrm{ml}}/(1-R)),
\label{ode_rec}
\end{equation}

\noindent where the rate of return of gas from a reservoir of ejected gas of mass $M_{\mathrm{res}}$ is given by 
$\dot{M}_{\mathrm{ret}} = M_{\mathrm{res}} / t_{\mathrm{ret}}$. For the case where $\beta_{\mathrm{ml}} \gg 1$,
Eqn.~\ref{ode_rec} simplifies to 

\begin{equation}
f_{\mathrm{b}} \dot{M}_{\mathrm{H}} + \dot{M}_{\mathrm{ret}} \approx \dot{M}_\star \frac{\beta_{\mathrm{ml}}}{1-R} = \dot{M}_{\mathrm{ej}}.
\label{ode_recbeta}
\end{equation}

\noindent The gas mass in the reservoir is given by

\begin{equation}
M_{\mathrm{res}} = \int_{0}^{t_{\mathrm{age}}} \left(\dot{M}_{\mathrm{ej}} -  \dot{M}_{\mathrm{ret}} \right) \, \mathrm{d}t.
\label{mres_integral}
\end{equation}

\noindent For the case where the halo mass accretion rate is approximately constant over a time scale, $t_{\mathrm{age}}$, substituting 
Eqn.~\ref{ode_recbeta} into \ref{mres_integral} yields

\begin{equation}
M_{\mathrm{res}} \approx \int_{0}^{t_{\mathrm{age}}} f_{\mathrm{b}} \dot{M}_{\mathrm{H}} \, \mathrm{d}t \approx f_{\mathrm{b}} \dot{M}_{\mathrm{H}} \, t_{\mathrm{age}}.
\label{mres_integral2}
\end{equation}

\noindent Therefore, in this idealised case $\dot{M}_{\mathrm{ret}}$ can be written as

\begin{equation}
\dot{M}_{\mathrm{ret}} = M_{\mathrm{res}} / t_{\mathrm{ret}} \approx f_{\mathrm{b}} \dot{M}_{\mathrm{H}} \, \frac{t_{\mathrm{age}}}{t_{\mathrm{ret}}}.
\label{reincorp_rate}
\end{equation}

\noindent Combining Eqns \ref{reincorp_rate} and \ref{ode_recbeta}, we find that

\begin{equation}
f_{\mathrm{b}} \dot{M}_{\mathrm{H}} \left(1 + \frac{t_{\mathrm{age}}}{t_{\mathrm{ret}}}\right) \approx \dot{M}_\star \frac{\beta_{\mathrm{ml}}}{1-R}.
\label{ode_recbeta2}
\end{equation}

\noindent In \galform, the return timescale is parametrised as

\begin{equation}
t_{\mathrm{ret}} = \frac{t_{\mathrm{dyn}}}{\alpha_{\mathrm{reheat}}} = \frac{1}{\alpha_{\mathrm{reheat}}} \sqrt{\frac{3}{4 \pi G \bar{\rho}_{\mathrm{H}}}},
\label{tret}
\end{equation}

\noindent where $t_{\mathrm{dyn}} = R_{\mathrm{vir}} / V_{\mathrm{vir}}$ is the dynamical timescale of the halo and $\alpha_{\mathrm{reheat}}$ is a dimensionless
model parameter set to $1.26$ for our fiducial model. As before, we can use Eqn.~\ref{collapse} and
adopt the case of an $\Omega_{\mathrm{M}} = 1$ universe, yielding

\begin{equation}
t_{\mathrm{ret}} = \frac{t_{\mathrm{age}}}{2 \pi \alpha_{\mathrm{reheat}}}.
\label{tret2}
\end{equation}

\noindent Examination of the bottom panel of Fig.~\ref{timescales} shows that this is a reasonable approximation.
Finally, combining eqns \ref{tret2} and \ref{ode_recbeta2} yields

\begin{equation}
f_{\mathrm{b}} \dot{M}_{\mathrm{H}} (1 + 2 \pi \alpha_{\mathrm{reheat}}) \approx \dot{M}_\star \frac{\beta_{\mathrm{ml}}}{1-R}.
\label{ode_recbeta3}
\end{equation}

\noindent Therefore, for the idealised case where $\beta_{\mathrm{ml}} \gg 1$, $t_{\mathrm{cool}} < t_{\mathrm{ret}}$, and $\dot{M}_{\mathrm{H}}$ remaining 
approximately constant over a time scale, $t_{\mathrm{age}}$, then the effect of including gas recycling is to increase the 
amount of gas available for star formation roughly by a factor of $1 + 2 \pi \alpha_{\mathrm{reheat}}$. For our fiducial \galform model with $\alpha_{\mathrm{reheat}} = 1.26$, this factor is 
$\sim 9$. We note that this modulation factor is completely independent of galaxy stellar mass, provided $\beta_{\mathrm{ml}} \gg 1$. 
Finally, as Equation~\ref{ode_recbeta3} is equivalent to Equation~\ref{ode_sfr} multiplied by
a constant factor, repeating the exercise of integrating Equation~\ref{ode_recbeta3} will give the same result that the specific stellar mass
assembly rate is simply given by $\dot{M}_\star / M_\star = 1/t_{\mathrm{age}}$, provided that $V_{\mathrm{vir}}$ remains constant with time.

\section{Towards Reproducing The Inferred Stellar Mass Assembly Histories Of Star-forming Galaxies}
\label{Modifications_Section}

In Section~\ref{SFH_Section} and Appendix~\ref{Invariance_Section}, we have demonstrated that for the standard parametrisations of 
supernova feedback, star formation and gas reincorporation used in \galform, it is not possible to 
reproduce the stellar mass assembly histories of star forming galaxies inferred from observations. 
The next logical step is to consider how these parametrisations would need to be changed in order to 
better reproduce the inferred observational trends. Clearly, the ideal scenario is to change the 
parametrisations such that they are more physically motivated and/or satisfy direct empirical constraints. 
The opposite and less desirable extreme is to use increasingly flexible parametrisations which have to be 
constrained statistically to reproduce global diagnostics of the galaxy population. Currently, the 
implementation of star formation in \galform can be argued to fall into the former case while the default 
implementations of feedback and reincorporation fall into the latter. We therefore choose to focus on how the 
implementation of feedback and gas reincorporation could be modified to change the model predictions relevant 
to our analysis.

\subsection{Modifying the mass loading factor for supernova feedback}

From the comparison between the predicted and inferred efficiency of stellar mass assembly shown in 
Fig.~\ref{sfe}, it is clear that the degree of coevolution between stellar mass and halo mass 
assembly rates needs to be reduced in \galform in order to reproduce the trends we infer according to the 
observations. This requirement appears to be particularly pertinent from intermediate through to late
times (roughly in the redshift range, $0 < z < 1$), where the efficiency of converting accreted gas into 
stars is inferred from the observations to drop after a peak at intermediate redshift. In Section~\ref{Explain_SFH_Section}, we 
demonstrated that the efficiency of feedback in our fiducial model, characterised by the mass loading factor, 
$\beta_{\mathrm{ml}}$, does not vary significantly over this redshift range. This is because the disk circular 
velocity does not evolve strongly over the lifetime of a typical star forming galaxy in \galform. A different
parametrisation for $\beta_{\mathrm{ml}}$ that does not depend only on circular velocity could potentially change this 
behaviour.

\cite{Lagos13} have recently introduced an alternative parametrisation for the ejection of gas from
galaxy disks and bulges as a result of feedback from supernovae \cite[see also ][]{Creasey13}. This offers a natural starting point
for our investigation because their work is motivated on physical grounds. Briefly, their methodology
is to track the evolution of bubbles driven by supernovae as they expand into the ambient ISM. They calculate the rate at which mass entrained
in these bubbles escapes vertically out of the disk and find that $\beta_{\mathrm{ml}}$ cannot be naturally parametrised
as a function of the disk circular velocity. Instead, they find that $\beta_{\mathrm{ml}}$ is better described as a
function of the gas fraction in the disk, $f_{\mathrm{g}}$, and either the total gas surface density, $\Sigma_{\mathrm{g}}$,
of the disk at the half mass radius, $r_{\mathrm{50}}$, or the gas scaleheight, $h_{\mathrm{g}}$, of 
the disk at the half mass radius. From this point onwards we refer to the former as the surface density
parametrisation and the latter as the scaleheight parametrisation. The surface density parametrisation is 
given by

\begin{equation}
\beta_{\mathrm{ml}} = \left[\frac{\Sigma_{\mathrm{g}}(r_{\mathrm{50}})}{1600 \mathrm{M_\odot pc^{-2}}}\right]^{-0.6} \left[\frac{f_\mathrm{gas}}{0.12}\right]^{0.8},
\label{beta_sdensity}
\end{equation}

\noindent and the scaleheight parametrisation is given by 

\begin{equation}
\beta_{\mathrm{ml}} = \left[\frac{h_{\mathrm{g}}(r_{\mathrm{50}})}{15 \mathrm{pc}}\right]^{1.1} \left[\frac{f_\mathrm{gas}}{0.02}\right]^{0.4}.
\label{beta_scaleheight}
\end{equation}

\begin{figure}
\includegraphics[width=20pc]{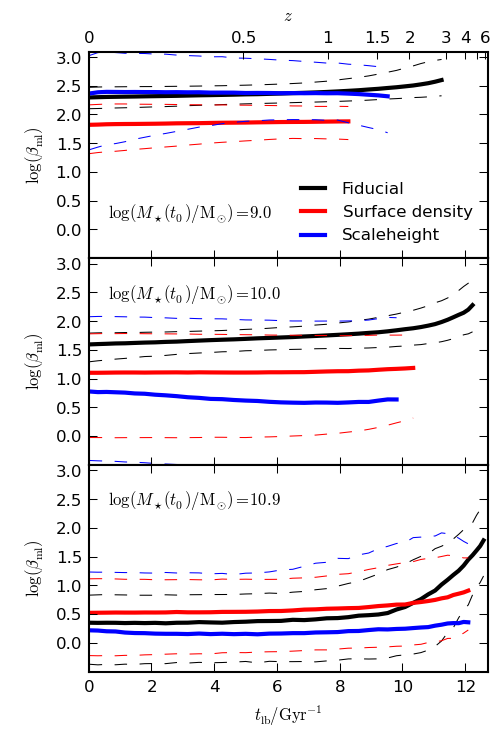}
\caption{The average evolution in the mass loading factor of outflows, $\beta_{\mathrm{ml}}$, for model galaxies which are central and star forming at $z=0$.
Model galaxies are binned according to their $z=0$ stellar mass, with each panel corresponding to a different mass bin.
The median $z=0$ stellar mass in each bin is labelled in each panel.
Solid black lines show the mean mass loading factor from our fiducial \galform model.
Dashed black lines show the corresponding median, $10^{\mathrm{th}}$ and $90^{\mathrm{th}}$ percentiles.
Red lines show the same information but for a version of our fiducial model modified to use the surface density mass loading parametrisation, given by \protect Eqn.~\ref{beta_sdensity} \protect\citep{Lagos13}.
Blue lines show the same information but for a version of our fiducial model modified to use the scaleheight mass loading parametrisation, given by \protect Eqn.~\ref{beta_scaleheight}.}
\label{beta_modified}
\end{figure}

We have used both of these parametrisations as separate modifications to our fiducial model
and find that neither significantly changes the shapes of the stellar mass assembly histories
of star forming galaxies to the extent that the model predictions are brought into better agreement with the data.
The reason for this failure is illustrated in Fig~\ref{beta_modified}, where we compare the average 
evolution in $\beta_{\mathrm{ml}}$ for galaxies that are star forming and central at $z=0$ between the different
models. It can be seen that although the modifications change the overall normalisation of $\beta_{\mathrm{ml}}$
and the dependence on $M_\star(t_0)$, the modified models actually result in even less evolution
of $\beta_{\mathrm{ml}}$ over the lifetime of a typical star forming galaxy. Further investigation shows that
this outcome arises because the effect on $\beta_{\mathrm{ml}}$ caused by the decline in the surface densities of 
star forming galaxies as they evolve is cancelled out by a corresponding drop in the gas fractions.

\subsection{Modifying the gas reincorporation timescale}
\label{tret_mod_section}

In addition to the mass loading factor, $\beta_{\mathrm{ml}}$, the way that ejected gas is 
reincorporated back into haloes is also modelled in a phenomenological manner in \galform.
Therefore, an alternative to modifying the mass loading factor supernova feedback in our model is to alter
the timescale for gas reincorporation for gas that has been ejected, $t_{\mathrm{ret}}$. At this stage, 
we choose to revert to the default parametrisation of $\beta_{\mathrm{ml}}$ (which depends on circular 
velocity) at this stage because using the modified models from \cite{Lagos13} would require substantial retuning of various 
model parameters to recover an agreement with the observed local luminosity and stellar mass functions.

In the third panel of Fig.~\ref{timescales}, we show the ratio of the characteristic gas 
reincorporation timescale relative to the age of the universe as a function of lookback time for our fiducial model.
As discussed in Section~\ref{Explain_SFH_Section}, this ratio of timescales evolves very little 
over the lifetime of a typical star forming galaxy, partly explaining the close levels of 
coevolution between stellar and halo mass assembly seen in Fig.~\ref{sfe} for our fiducial model.
The most desirable step at this stage would be to formulate a physically motivated model for
gas reincorporation timescales in the hope that a more realistic model could change this 
behaviour. Such an undertaking is beyond the scope of this study, but as an intermediate step,
we instead introduce an {\it ad hoc} modification to the parametrisation of gas reincorporation 
timescales used in \galform. We note that this step essentially amounts to an empirical fit to 
the trends which we infer from the data and is of little scientific value in itself. However, the 
resulting evolution in the reincorporation timescale for star forming galaxies can serve as a 
guide for the development of a physically motivated model in future work.

To match the shape of the stellar mass assembly histories inferred from the data shown
in Fig.~\ref{hfh_sfh_comp}, we consider a model where $t_{\mathrm{age}} / t_{\mathrm{ret}}$ rises
from early times to a peak at $z=2$, before falling to $z=0$. A natural way to achieve an
early time rise is to make $t_{\mathrm{age}} / t_{\mathrm{ret}}$ correlate positively with halo 
mass. This is the same scaling that \cite{Henriques13} adopt in order to allow their
model to reproduce the observed evolution in the stellar mass function.
However, a natural scaling that results in a drop in $t_{\mathrm{age}} / t_{\mathrm{ret}}$ at late
times is less obvious and we instead choose to simply introduce an arbitrary function of redshift to
achieve this. After a process of experimentation and iteration, we arrive at the following
modified parametrisation for gas reincorporation,

\begin{equation}
\dot{M}_{\mathrm{hot}} = \frac{\alpha_{\mathrm{reheat}}}{t_{\mathrm{dyn}}} \, \left(\frac{M_{\mathrm{H}}}{10^{11.9}\mathrm{M_\odot}}\right) \, f(z),
\label{reincorporation_modified}
\end{equation}

\noindent where $f(z)$ is given by

\begin{equation}
\log[f(z)] = 6 \, \exp{\left[-\frac{(1+z)}{3}\right]} \, \log[1+z].
\label{fudge_function}
\end{equation}

At this stage we note that in Appendix~\ref{Invariance_Section}, we show that the 
shapes of the stellar mass assembly histories predicted by \galform are almost invariant under 
changes in the model parameters which control the relationship between stellar and halo mass. In
other words, this means that until now, our results for the shape of the stellar mass assembly 
histories of star forming galaxies have been independent of whether or not the model provides a 
good match to the $z=0$ stellar mass function. However, once we change the parametrisation of the
gas reincorporation in \galform, this feature of the model may not be preserved. Consequently, 
we now have to consider whether our modified \galform models can also reproduce the
$z=0$ stellar mass function, particularly because we have introduced a dependence on halo mass
into Eqn.~\ref{reincorporation_modified}. We find that we can recover reasonable agreement
with the local stellar mass function simply by fine tuning the various model parameters that
appear in Eqn.~\ref{reincorporation_modified}. We also reduce the threshold for AGN feedback 
to be effective at suppressing gas cooling in haloes by changing the model
parameter $\alpha_{\mathrm{cool}}$ from $1.0$ to $1.3$ \cite[see ][]{Bower06}.
From this point onwards, we refer to this modified model simply as the modified reincorporation model.

\begin{figure}
\includegraphics[width=20pc]{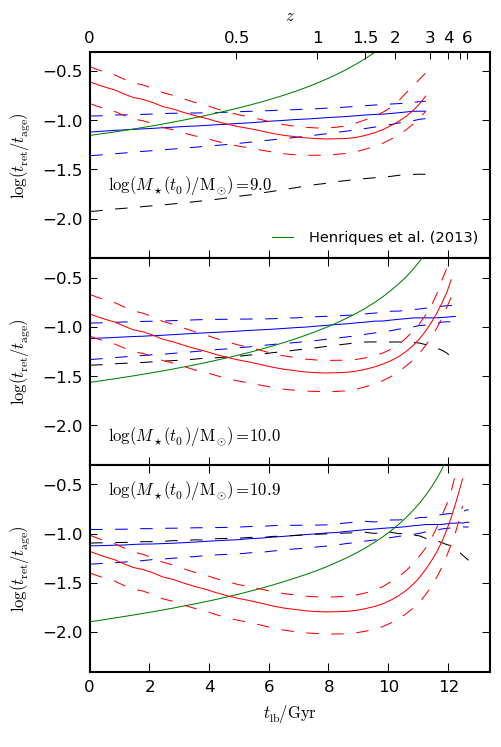}
\caption{The ratio of the average reincorporation timescale to the age of the universe for model galaxies which are central and star forming at $z=0$, plotted as a function of lookback time.
Model galaxies are binned according to their $z=0$ stellar mass and each panel shows a different stellar mass bin.
The median $z=0$ stellar mass in each bin is labelled in each panel.
Solid blue lines show the medians of the distribution from our fiducial \galform model.
Dashed blue lines show the corresponding $10^{\mathrm{th}}$ and $90^{\mathrm{th}}$ percentiles.
Red lines show the same information but for the modified reincorporation model.
Green solid lines show the median of the distribution for galaxies from our fiducial model which would be obtained if we were to use the reincorporation timescale from eqn. $8$ in \protect\cite{Henriques13}.
For reference, black dashed lines show the ratio of the median cooling timescale, $t_{\mathrm{cool}}$, to the age of the universe for model galaxies from our fiducial model.}
\label{tret_modified}
\end{figure}

A comparison between our fiducial model and the modified reincorporation model for the evolution 
in $t_{\mathrm{ret}} / t_{\mathrm{age}}$ for star forming galaxies is presented in Fig.~\ref{tret_modified}.
In addition, Fig.~\ref{tret_modified} also shows the corresponding timescale proposed 
by \cite{Henriques13} which scales only with the virial mass.
By construction, $t_{\mathrm{ret}} / t_{\mathrm{age}}$ evolves much more strongly in our modified
reincorporation model than in our fiducial model. Additionally, the dispersion in $t_{\mathrm{ret}} / t_{\mathrm{age}}$ can be slightly 
larger in the modified model for some lookback times. Given that Eqn.~\ref{reincorporation_modified}
introduces a dependence on halo mass, the change is presumably caused by scatter in the relationship between 
the stellar mass and halo mass of central star forming galaxies. This is noteworthy because 
any change in the scatter in $t_{\mathrm{ret}} / t_{\mathrm{age}}$ could have an effect on the scatter of
the star forming sequence predicted by our modified reincorporation model.

The evolution in the cooling timescale is also shown in Fig.~\ref{tret_modified}. It can be seen
that for our modified reincorporation model, the cooling timescale can become significantly longer than
the reincorporation timescale for the most star forming galaxies. In this regime, gas that is rapidly
reincorporated back into the halo will be delayed from returning to the galaxy disk until the gas is
able to cool. This is important because further shortening the reincorporation timescale in this regime 
will cease to have a significant impact on the rate at which gas is made available for star formation.

\begin{figure*}
\begin{center}
\includegraphics[width=40pc]{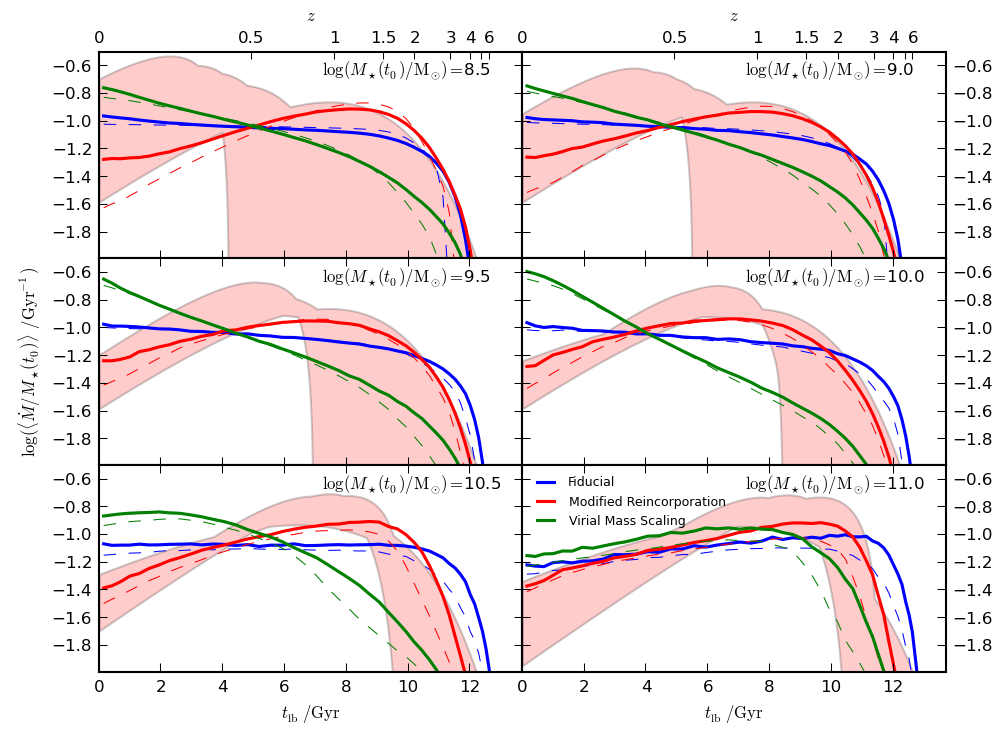}
\caption{The average stellar mass assembly histories of galaxies that are star forming at $z=0$, plotted as a function of lookback time.
Blue solid lines show predictions from our fiducial \galform model for the mean mass assembly histories of the main stellar progenitors of central galaxies.
Dashed blue lines show the corresponding medians of the distribution.
Red lines show the same information but for the modified reincorporation model.
Green lines show the same information but for a model using the virial mass scaling for the reincorporation timescale proposed by \protect\cite{Henriques13}.
Model galaxies are binned according to their $z=0$ stellar mass with each panel corresponding to a different mass bin.
The median $z=0$ stellar mass in each bin is labelled in each panel.
The filled pink region shows the range of stellar mass assembly histories that are inferred by applying the MSI technique to observational data from the literature.}
\label{sfh_modified}
\end{center}
\end{figure*}

A comparison between our fiducial model and the modified reincorporation model for the predicted stellar
mass assembly histories of star forming galaxies is presented in Fig.~\ref{sfh_modified}. Again, by 
construction we have tuned the modified reincorporation model in order to ensure qualitative agreement
with the pink shaded region inferred from the observations using MSI. Comparison with the stellar mass
assembly histories inferred from observational data shown in Fig.~\ref{sfh_obs}
shows that this agreement holds with MSI applied to observational data where the slope of the star 
forming sequence, $\beta_{\mathrm{sf}} \approx 0$. As discussed in Section~\ref{sfh_model_obs_comp}, a lower value of 
$\beta_{\mathrm{sf}}$ introduces a strong downsizing trend into the stellar mass assembly histories of star forming
galaxies which is difficult to reconcile with the approximately self-similar shape of the halo
mass assembly histories predicted by the $\Lambda$CDM cosmological model. In principle, we could adjust 
Eqn.~\ref{reincorporation_modified} even further to try to reproduce this downsizing trend. However,
we have already introduced a very strong redshift scaling into the reincorporation timescale. Therefore, 
we choose to present a modified model which is closest to our fiducial model while still showing consistency 
with the pink shaded region in Fig.~\ref{sfh_modified}.

Finally, we show the comparison between our fiducial model and the modified reincorporation model for the
evolution in the specific star formation rates of star forming galaxies in Fig.~\ref{ssfr_zm_Lagos_obs_comp}.
Our modification to the reincorporation timescale has mixed success. For the top two panels, corresponding 
to the $\log(M_\star / \mathrm{M_\odot}) = 9.5,10$ bins, the modified model shows a significantly improved 
agreement with the observational trend. Unlike for the fiducial model, the evolution in the mean specific 
star formation rates in the modified model does not trace the inverse of the age of the Universe as a 
function of lookback time. Instead, specific star formation rates are elevated at early times before dropping 
below the fiducial model at $z < 0.5$. For the $\log(M_\star / \mathrm{M_\odot}) = 10.5$ bin, the modified 
reincorporation model has a steeper drop below $z \approx 0.5$ compared to the fiducial model but the two 
models are very similar at higher redshifts, in disagreement with the data. The two models are very similar 
for all lookback times in the $\log(M_\star / \mathrm{M_\odot}) = 11$ bin and are both in disagreement with 
the data. 

At first glance it is puzzling that modified reincorporation model fails to reconcile the model predictions
with the observational data for massive star forming galaxies while it does qualitatively reproduce the inferred
stellar mass assembly histories shown in Fig.~\ref{sfh_modified}. However, it must be kept in mind that our modification 
to the reincorporation timescale was constructed only to reproduce the shape of the stellar mass assembly 
histories of galaxies that are still star forming at $z=0$.
Galaxies observed at $z > 0$ in the most massive stellar mass bins shown in Fig.~\ref{ssfr_zm_Lagos_obs_comp} will,
typically, have dropped below the star forming sequence by $z=0$. Therefore, the specific star formation rates
of the most massive galaxies at high redshift will not have been constrained by our analysis of stellar mass assembly
histories of galaxies that are still star forming at $z=0$. Furthermore, galaxies that are quenched above
$z \approx 1-2$ will be less affected by the rapid evolution in the reincorporation timescale which we impose in 
Eqn.~\ref{fudge_function} below $z=2$. This highlights the need for a physical model of gas 
reincorporation rather than the artificial redshift scaling which we use here. In addition, for the most massive
star forming galaxies, we show in Fig.~\ref{tret_modified} that the cooling timescales become long relative to the
modified reincorporation timescales. As discussed earlier, this will reduce the impact of any modification towards
shorter reincorporation timescales. Finally, the mass loading factor, $\beta_{\mathrm{ml}}$, for the most massive galaxies 
is smaller than for lower mass galaxies. Therefore, a larger fraction of gas accreted onto the haloes of these systems
for the first time will be able to form stars without being affected by the reincorporation timescale.

Fig.~\ref{sfh_modified} and Fig.~\ref{ssfr_zm_Lagos_obs_comp} also show the stellar mass
assembly histories and specific star formation rate evolution predicted by an alternative 
\galform model that uses the virial mass scaling for the reincorporation timescale proposed 
by \protect{\cite{Henriques13}}. Details of this model and a discussion of the differences 
with our modified reincorporation model are presented in Appendix~\ref{H12_section}. To summarise, we
find that this alternative model with the virial mass scaling is considerably more successful 
than either our fiducial model or our modified reincorporation model in reproducing the 
observed evolution in the stellar mass function. However, Fig.~\ref{sfh_modified}
and Fig.~\ref{ssfr_zm_Lagos_obs_comp} show that this alternative model fails to reproduce the 
stellar mass assembly and star formation rate evolution inferred from observations. 
We strongly emphasise that this result will not necessarily hold for the \cite{Henriques13} 
model where the treatment of gas that is ejected from galaxy disks differs from the 
\galform model.

\section{Discussion}
\label{Discussion_Section}

The focus of this study has been on using the observed evolution of the star forming sequence
as a constraint on galaxy formation models. The disagreement in this evolution between models and observational data is undoubtedly
related to the problems with reproducing the correct evolution in the low mass end of the stellar mass function which
has recently received considerable attention in the literature \cite[e.g.][]{Avila-Reese11,Weinmann12,Henriques13,Lu13a,Lu13b}. 
Specifically, there is a general finding that models and simulations overpredict the ages of low mass galaxies and consequently underpredict 
evolution in the low mass end of the stellar mass function at low redshift. \cite{Weinmann12} interpret this problem as an indication that the level
of coevolution between halo and stellar mass assembly needs to be reduced, broadly in agreement with our results.
However, part of the reason why they arrive at this conclusion is because they identify the 
prediction of a positive correlation between specific star formation rate and stellar mass as a key problem with 
respect to the data. We note that in contrast, \galform naturally predicts a slightly negative correlation for star forming 
galaxies and that this is also true for many other models and simulations presented in the literature \cite[e.g.][]{Santini09,Dutton10,Lamastra13,Torrey14}.

\cite{Henriques13} show that there is no combination of parameters for their standard galaxy formation model that can
reconcile the model with the observed evolution in the stellar mass and luminosity functions. This is consistent
with the findings of \cite{Lu13a}, who use a similar methodology but for a different model. \cite{Lu13b} compare three different
models of galaxy formation and find that they all predict very similar stellar mass assembly histories and suffer from
predicting too much star formation at high redshift in low mass haloes. We note that the models presented in \cite{Lu13b} 
are all very similar to \galform in many respects and that therefore the similarity of the predictions from their three 
models makes sense in the context of the discussion we present in Appendix~\ref{Invariance_Section}. 

\cite{Henriques13} go one step further to suggest an empirical modification to the reincorporation timescale within their 
model that reduces the rate of star formation at early times in low mass haloes. In this respect, their equation $8$ uses
the same scaling between reincorporation timescale and halo mass which we introduce in Eqn.~\ref{reincorporation_modified} for the same reason. 
However, our modification diverges from their suggestion in that we also require an additional redshift dependence that
lengthens the reincorporation timescale towards low redshift. The modification suggested by \cite{Henriques13} can be 
compared to our modification in Fig.~\ref{tret_modified}.
The difference between the two suggested modifications stems from the way that our analysis indicates that it is not simply that
stars form too early in the model. Instead, we find that it is the precise shape of the stellar mass assembly history which is inconsistent with the
currently available data which favours a peak of activity at intermediate times. This highlights how the differences in methodology 
between different studies can lead to different conclusions. Our analysis is designed to reduce the number of relevant physical 
processes by focusing only on the normalisation of the star forming sequence. In principle, this approach can provide a more direct insight into 
how the implementation of different physical processes within galaxy formation models needs to be changed, provided that the uncertainty 
in the relevant observations can be correctly accounted for. On the other hand, as discussed in Appendix~\ref{H12_section}, our modified reincorporation
model does not reproduce the evolution in the stellar mass function inferred from recent observations. We again emphasise that the focus of this study 
is on the evolution of the normalisation of the star forming sequence and that the stellar mass function can be affected by the quenching processes which
we have not considered in our analysis. Nonetheless, it may well be 
the case that our methodology is limited by the lack of a consensus on the slope of the star forming sequence in observations. Alternatively, there could be
some inconsistency between observations of the star forming sequence and observations of the evolution in the stellar mass function. We note that the latter 
possibility is disfavoured by recent abundance matching results \cite[e.g.][]{Behroozi13,Moster13}.

\subsection{Do the stellar mass assembly histories of star forming galaxies rise and then fall?}

Our suggestion that the reincorporation timescale needs to be increased at low redshift stems from our 
inference from observations that the stellar mass assembly histories of star forming galaxies rise to a peak 
before falling towards the present day. As discussed in Appendix~\ref{MSI_Obs_Section}, this inference is 
consistent with the findings of \cite{Leitner12} who use a similar methodology, albeit with the caveat 
that we find that evidence of a strong downsizing trend in the purely star forming population is not 
conclusive. Instead, we find that the considerable uncertainty that remains in the power-law slope of 
the star forming sequence means that overall, the observational data are also consistent with no downsizing,
such that the shapes of the stellar mass assembly histories of star forming galaxies are independent of the final stellar
mass.
Clearly, any improvements in measuring the form of the star forming sequence as  
a function of lookback time would greatly increase the constraining power of the MSI technique with respect to galaxy formation
models. If the slope of the sequence, $\beta_{\mathrm{sf}}$, can be conclusively shown to be significantly 
below zero as advocated, for example by \cite{Karim11}, then even larger modifications than those considered 
here towards separating stellar and halo mass assembly would be required.

Another methodology that can be used to infer the shape of the stellar mass assembly histories of galaxies is to 
employ abundance matching to make an empirical link between the dark matter halo population predicted by theory and the 
observed galaxy population \cite[e.g.][]{Behroozi13,Moster13,Yang13}. Comparison with stellar mass assembly histories of the star forming galaxies that are
discussed in this study is complicated by the fact that abundance matching has only been used so far to predict the average star
formation histories of all galaxies (including passive galaxies) and as a function of halo mass. On average, the haloes
hosting the galaxies which we consider in this study have median masses of $\log(M_{\mathrm{H}} / \mathrm{M_\odot}) < 12$, where 
the fraction of passive central galaxies relative to star forming centrals is predicted to be negligible. However, because
there is substantial scatter between stellar mass and halo mass for central galaxies, the fraction of passive galaxies
at a given stellar mass is not negligible for most of the stellar mass bins which we consider in this study. For example,
the fraction of central galaxies with $\log(M_\star / \mathrm{M_\odot}) = 10$ that are passive is predicted to be 
$25 \%$ at $z=0$ in our fiducial \galform model. Furthermore, the star forming galaxies considered in this study and in 
\cite{Leitner12} are hosted by haloes that reside within a fairly narrow range of halo mass. If we ignore these issues, 
then qualitatively speaking, it is apparent that the shape of stellar mass assembly histories inferred by
\cite{Behroozi13} and \cite{Yang13} are broadly consistent with what we and \cite{Leitner12} infer from the data, in that 
there is a rise with time towards a peak at some intermediate redshift before a fall towards the present day. \cite{Moster13}
show qualitative agreement with this picture for $\log(M_{\mathrm{H}} / \mathrm{M_\odot}) = 12$ haloes, but find a constant rise from early to late times in the stellar 
mass assembly rates of galaxies that reside within haloes with $\log(M_{\mathrm{H}} / \mathrm{M_\odot}) = 11$. 

Finally, we also note that \cite{Pacifici13} find that the spectral energy distributions (SEDs) of massive star forming galaxies are well
described by models that feature initially rising then declining star formation histories. However, for lower mass galaxies they find that the
SEDs are best reproduced using star formation histories that monotonically rise towards the present day, in qualitative agreement
with the results from \cite{Moster13}. However, their galaxy sample does not include any galaxies observed below $z=0.2$, corresponding
to a lookback time of $t_{\mathrm{lb}} \approx 3 \, \mathrm{Gyr}$. It is therefore unclear whether their analysis disfavours a drop
in the star formation rates of lower mass galaxies at late times.

\subsection{Modifications to galaxy formation models}
 
The parametrisations for star formation and feedback that are implemented in most galaxy formation models can reproduce 
the shape of the local luminosity and stellar mass functions. However, as observational
data that characterises the evolution of the galaxy population has improved, it has now been demonstrated that either one or 
more of these parametrisations is inadequate or alternatively that another important physical process has been neglected in 
the models entirely. 
The assumption that the reincorporation timescale for ejected gas scales with the dynamical timescale of the host halo is common
to various semi-analytic galaxy formation models \citep[e.g.][]{Bower06,Croton06,Somerville08,Lu11}. If the reincorporation timescale
is set to exactly the dynamical timescale, the associated physical assumption is that ejected gas simply behaves in a ballistic manner,
ignoring any possible hydrodynamical interaction between the ejected gas and the larger scale environment.
In practice, these models (including ours) typically introduce a model parameter such that the reincorporation timescale is not exactly equal to
the dynamical timescale, reflecting the considerable uncertainty on predicting this timescale. Nonetheless, the assumption that this
uncertainty can be represented by a single parameter and that there is no additional scaling with other galaxy or halo properties is
clearly naive. Comparison with hydrodynamical simulations will clearly be useful in this respect, provided that the reincorporation rates
can be clearly defined and measured from the simulations and that the effect of the assumptions made in sub-grid feedback models can be
understood.

While we and \cite{Henriques13} show that a modification to the 
reincorporation timescale for gas ejected by feedback can be one solution, we could equally
change the parametrisation for the mass loading factor, $\beta_{\mathrm{ml}}$, or the star formation law introduced in \cite{Lagos11a}.
In this analysis, we found that the physically motivated parametrisation for the mass loading factor of SNe driven winds presented 
in \cite{Lagos13} fails to reconcile the model with the data. However, it should be noted that unlike the fiducial model we 
consider for this study, the supernova feedback model presented in \cite{Lagos13} relies heavily upon correctly predicting the evolution
in the sizes of galaxies. In principle, if the predicted sizes evolved differently in our model, it is possible that using the \cite{Lagos13} supernova
feedback model could help to reconcile model predictions for the stellar mass assembly histories of galaxies with the observational data.
As for modifying the star formation law, the implementation used in \galform is derived from direct empirical constraints. Furthermore, changing 
the star formation law will have little impact on the stellar mass assembly histories of star forming galaxies as long as the 
characteristic halo accretion timescale is longer than the disk depletion timescale.
Of course, an alternative to the physically motivated \cite{Lagos13} model is simply to implement an ad hoc modification to the mass loading, similar to that
given by Eqn.~\ref{reincorporation_modified} for the reincorporation timescale. We note that by doing this, we find it is possible to produce a model that almost exactly matches
the predictions made by the modified reincorporation model presented in this paper. It therefore suffers from the same problems as the modified
reincorporation model in reproducing the observed evolution of the stellar mass function and the decline in the specific star formation rates
of the most massive star forming galaxies at a given redshift.

Many other suggestions for changing the stellar mass assembly histories predicted by models and simulations have been made recently in the literature, typically focusing on 
reducing the fraction of stars that form at high redshift. For example, \cite{Krumholz12} argue that early star formation
is reduced once the dependence of star formation on metallicity is properly implemented in hydrodynamical simulations.
\cite{Gabor14} suggest that if galaxies at high redshift accrete directly from cold streams of gas, the accreted gas injects 
turbulent energy into galaxy disks, increasing the vertical scaleheight and consequently lowering the star formation efficiency
in these systems by factors of up to $3$. \cite{Lu14} demonstrate that if the circum-halo medium can be preheated at early times
up to a certain entropy level, the accretion of baryons onto haloes can be delayed, reducing the amount of early star formation. 
Various authors \cite[e.g][]{Aumer13,Stinson13,Trujillo-Gomez13} find that implementing a coupling between the radiation emitted by young stars and the
surrounding gas into their simulations can significantly reduce the levels of star formation in high redshift galaxies. 
\cite{Hopkins13a} and \cite{Hopkins13b} echo these findings and emphasise the highly non-linear nature of the problem once sufficient resolution is obtained to start resolving 
giant molecular cloud structures. They argue that radiative feedback is essential to disrupt dense star forming gas before SNe 
feedback comes into effect to heat and inject momentum into lower density gas, avoiding the overcooling problem as a result.
It remains to be seen at this stage whether the emergent behaviour from such simulations, once averaged over an entire galaxy disk or bulge, can be captured in the parametrisations
that are used in semi-analytic galaxy formation models.

\section{Summary}
\label{Summary_Section}

We have performed a detailed comparison between predictions from the \galform semi-analytic model of galaxy formation with 
observational data that describe the average star formation rates of star forming galaxies as a function of stellar mass 
and lookback time. To better understand the origin of discrepancies between the model and the data, we also use the 
observational data to infer the shape of the stellar mass assembly histories of galaxies that are still central and star forming
at the present day. This is achieved by integrating the inferred relationship between star formation rate and stellar mass for
star forming galaxies back in time from the present day. Crucially, we account for the considerable uncertainty 
that remains in the literature regarding the slope of the power-law dependence of star formation rate on stellar mass.
We then attempt to explain our results by analysing the timescales of the various physical processes in the model which are
important for shaping the stellar mass assembly histories of star forming galaxies. 

Our main results are summarised as follows:

\begin{itemize}

\item For our fiducial model, there are qualitative differences with the observational data in the way that the average specific star formation rates 
of star forming galaxies evolve with time at a given stellar mass. The model predicts average specific star formation rates that evolve 
too slowly with lookback time, tracing the inverse of the age of the universe at a given epoch. In contrast the observational data implies
that the average specific star formation rates of star forming galaxies grow exponentially as a function of lookback time. Quantitatively, this 
leads to discrepancies in the predicted average specific star formation rates of up to $0.5 \, \mathrm{dex}$ compared to the data.

\item We show that the main sequence integration technique, as advocated by \cite{Leitner12}, can qualitatively recover the shape of 
the stellar mass assembly histories of galaxies that are still star forming at the present day when it is applied to our fiducial model.

\item After applying this technique to a compilation of observational data, we show that there is a qualitative difference between
the inferred shape of the stellar mass assembly histories of star forming galaxies and the predictions from our fiducial model. Specifically,
the model predicts stellar mass assembly histories that are almost flat over most of the lifetime of star forming galaxies. In contrast,
the trend we infer from the data is that stellar mass assembly histories rise from early times, peak at an intermediate redshift and subsequently
fall towards the present day.

\item The exact position of the peak in these inferred stellar mass assembly histories depends sensitively on the slope of the star forming
sequence of galaxies. We show that no clear consensus on this slope has emerged yet from observations presented in the literature. For the case where the specific star 
formation rate is independent of stellar mass, the resulting shape of the stellar mass assembly histories of galaxies that are still star 
forming at the present day is also independent of stellar mass. For the case where there is a strong anti-correlation between specific
star formation rate and stellar mass, there is also a strong downsizing trend that emerges for this population of galaxies. In this case, 
less massive galaxies start forming stars at a later time with respect to more massive star forming galaxies. We emphasise that this
should be completely independent of processes that quench star formation in galaxies. Such a downsizing trend in the purely star 
forming population is difficult to reconcile with the approximately self-similar halo mass assembly histories predicted by simulations 
of structure formation.

\item The shapes of the stellar mass assembly histories predicted by our fiducial model are unaffected by changes to the various input
parameters to the \galform model. This is despite the fact that for the same changes to these model parameters, it is possible to 
significantly affect the present day relationship between stellar mass and halo mass.

\item The roughly flat stellar mass assembly histories predicted by our fiducial model arise because of the standard parametrisations
for supernova feedback that are implemented in semi-analytic galaxy formation models. The efficiency with which cold gas is ejected from 
galaxy disks evolves very little over the majority of the lifetimes of star forming galaxies. This comes as a result of the standard 
scheme used in semi-analytic models where the mass loading factor is a parametrised as a function of circular velocity which, in turn, is almost
constant over the lifetime of an individual star forming galaxy. Similarly, the timescale, relative to the age of the Universe, over which gas ejected by feedback is reincorporated
into galaxy haloes also varies very little for individual star forming galaxies. In this case, the typical assumption that the reincorporation 
timescale scales with the halo dynamical time results in this behaviour. We also show using simple arguments that when the efficiency of 
feedback does not vary with time for a given galaxy, the specific star formation rates of star forming galaxies will naturally trace the inverse 
of the age of the Universe at a given stellar mass.

We demonstrate that a modification to the reincorporation timescale, such that this timescale is lengthened at early and late
times, can produce peaked stellar mass assembly histories for galaxies that are still star forming at the present day. This modification significantly
improves the agreement with the data for the evolution in the average specific star formation rates of star forming galaxies with 
$9.5 < \log(M_\star / \mathrm{M_\odot}) < 10.0$. However, the modification is less effective for more massive star forming galaxies where
radiative cooling timescales become comparable to or longer than the corresponding reincorporation timescales. We also show that the
modification fails to reproduce the rapid evolution in the low mass end of the stellar mass function inferred from observations below $z=2$.

\end{itemize}

We conclude that modifications to the standard implementations of supernova feedback used in 
galaxy formation models and cosmological hydrodynamical simulations are probably required. Rather than altering the 
efficiency of feedback or star formation in a global sense over the lifetime of a given galaxy, it appears to be necessary 
to introduce a dependency that changes the efficiency of one or both of these processes with time.

\section*{Acknowledgements}

This work was supported by the Science and Technology Facilities Council [grant numbers ST/J501013/1, ST/F001166/1].
This work used the DiRAC Data Centric system at Durham University, operated by the Institute for Computational Cosmology on behalf of the STFC DiRAC HPC Facility (www.dirac.ac.uk). 
This equipment was funded by BIS National E-infrastructure capital grant ST/K00042X/1, STFC capital grant ST/H008519/1, and STFC DiRAC Operations grant ST/K003267/1 and Durham University. 
DiRAC is part of the National E-Infrastructure.
We thank Violeta Gonzalez-Perez for making suggestions which helped improve the clarity of this paper.

%----------------------------------------------
\bibliographystyle{mn2e}
\bibliography{bibliography}
%---------------------------------------------------------------------

\appendix

\section{Inferring The Stellar Mass Assembly Histories Of Star Forming Galaxies}
\label{MSI_Appendix}

This appendix introduces the main sequence integration (MSI) technique as a method of inferring the average stellar mass assembly
histories of star forming galaxies. We then present a discussion of testing the technique by attempting to recover the
stellar mass assembly histories predicted by \galform. Finally, details of the observational compilation used to infer the stellar mass
assembly histories of star forming galaxies and a discussion of the results is provided.

\subsection{Main sequence integration}
\label{MSI_Introduction}

The underlying idea of MSI is that an ``average'' galaxy can be tracked across the star
formation rate versus stellar mass plane by using measurements of the average star formation rate,
at a given stellar mass and lookback time, for galaxies which belong to a star forming sequence. 
This evolutionary track is then integrated, 
either forwards or backwards in time, from a specified starting mass, $M_\star(t_0)$, 
and starting time, $t_0$. For the case of integrating backwards in time, the resulting 
stellar mass assembly history is given by

\begin{align}
M_\star(t) &= M_\star(t_0) - \int_{t}^{t_0} \langle\psi(M_\star,t')\rangle \, \mathrm{d}t' \nonumber \\ &\qquad {} + \int_{0}^{t_0} \langle\psi(M_\star,t')\rangle \, R(t_0 - t') \, \mathrm{d}t'
\nonumber \\ &\qquad {} - \int_{0}^{t} \langle\psi(M_\star,t')\rangle \, R(t - t') \, \mathrm{d}t',
\label{msi}
\end{align}

\noindent where $\langle\psi(M_\star,t)\rangle$ is the average star formation rate of star forming 
galaxies of stellar mass $M_\star$ at time $t$ and $R(t)$ is the fraction of mass 
returned to the ISM by SNe and stellar winds for a simple stellar population of age $t$. 
For the case of integrating backwards in time, this equation can only be solved numerically using an 
iterative method in order to account for the returned fraction \cite[see][]{Leitner11}). 
For this study, we instead choose to be consistent with the approach used in \galform 
by adopting the instantaneous recycling approximation. In this case, $R(t)$ is replaced
by a constant (set to $0.39$ to be consistent with our fiducial \galform model) and 
Equation~\ref{msi} simplifies to

\begin{equation}
M_\star(t) = M_\star(t_0) - (1-R) \, \int_{t}^{t_0} \langle\psi(M_\star,t')\rangle \, \mathrm{d}t',
\label{msi_ira}
\end{equation}

\noindent which can be solved numerically using a simple Runge-Kutta integration 
scheme. The effect of assuming instantaneous recycling can be seen by examining Fig. 9 
in \cite{Leitner12}. Relative to the other uncertainties on the inferred stellar mass
assembly histories which we discuss later, we expect from their Fig. 9 that the 
effect of assuming instantaneous recycling is most likely negligible.

In order to calculate $M_\star(t)$ using Equation~\ref{msi_ira} at each timestep, the 
average star formation rate of star forming galaxies, $\langle\psi(M_\star,t)\rangle$, must be specified 
using measurements of the star forming sequence. Our parametrisation of 
$\langle\psi(M_\star,t)\rangle$ is described in Appendix~\ref{MSI_galform} for our application to 
\galform, and in Appendix~\ref{MSI_Obs_Section} for our application to a compilation of 
observational data. Finally, for a more intuitive link to the dark matter halo mass 
assembly histories which we consider later, we choose to work in terms of stellar mass 
assembly histories rather than star formation histories. As we assume instantaneous 
recycling, both in the model and when analysing the observational data, these are related trivially 
by linking the stellar mass assembly rate, $\dot{M}_\star$, to the star formation rate using
$\dot{M}_\star = (1-R) \, \psi(M_\star,t)$.

\subsection{Stellar mass assembly histories of \galform galaxies and validation of MSI}
\label{MSI_galform}

\begin{figure*}
\begin{center}
\includegraphics[width=40pc]{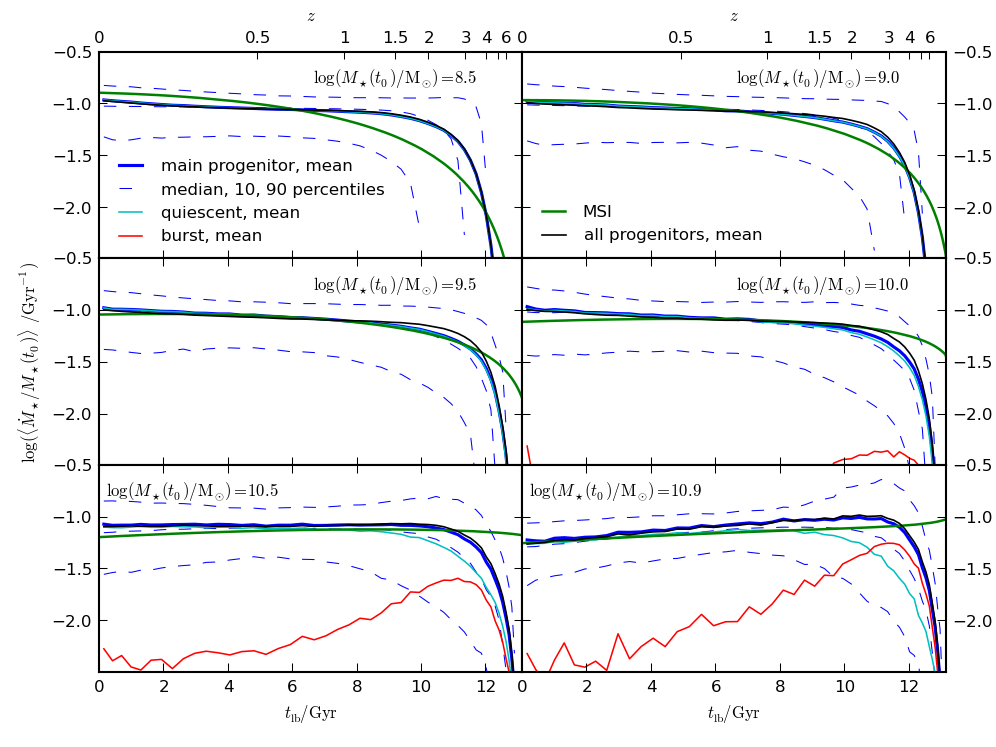}
\caption{The average stellar mass assembly histories from our fiducial \galform model of central galaxies that are star forming at $z=0$, plotted as a function of lookback time.
Model galaxies are binned by their stellar mass at $z=0$, with each panel corresponding to a different mass bin.
The median $z=0$ stellar mass in each bin is labelled in each panel.
Heavy blue solid lines show the mean stellar mass assembly histories for the main stellar progenitors, as calculated directly from our fiducial model.
Dashed blue lines show the corresponding $10^{\mathrm{th}}$, median and $90^{\mathrm{th}}$ percentiles of the distribution.
Cyan and red lines show, respectively, the contribution to the mass assembly histories from quiescent star formation and bursts.
Black lines show the mean stellar mass assembly histories but for the case of summing over all of the stellar progenitors of each $z=0$ galaxy.
These are largely coincident with the heavy blue lines.
Green lines show the stellar mass assembly histories calculated by applying the MSI technique to the star forming sequence predicted by our fiducial model.}
\label{sfh_Lagos}
\end{center}
\end{figure*}

In Fig.~\ref{sfh_Lagos} we show the average stellar mass assembly histories of galaxies from our 
fiducial \galform model that are central and star forming at $z=0$. 
The solid blue lines in Fig.~\ref{sfh_Lagos} show the mean stellar mass assembly histories taken
directly from our fiducial model. It can be seen that, roughly speaking, the overall shape of the 
mass assembly histories is nearly independent of the final stellar mass. Each stellar mass bin shows a 
sharp rise at early times before flattening out over the majority of the age of the universe. There is a 
slight deviation from this behaviour for galaxies with $M_\star(t_0) \approx 10^{11} \, \mathrm{M_\odot}$, 
which instead display a gradual decline in the stellar mass assembly rate after a peak at 
$t_{\mathrm{lb}} \approx 11 \, \mathrm{Gyr}$. The dashed blue lines in Fig.~\ref{sfh_Lagos} show 
the $10^{\mathrm{th}}$, median and $90^{\mathrm{th}}$ percentiles, indicating the spread in the distribution
around the mean.

As well as using the MSI technique to compare these model predictions with any trends inferred from 
observational data, it is useful to test how well the MSI technique works when applied to the
star forming sequence predicted by our fiducial model. To apply MSI to \galform, it is necessary to 
first specify the form of the star forming sequence in the model by parametrising $\langle\psi(M_\star,t)\rangle$.
In principle, we could tabulate this for all of the output times used to generate the assembly 
histories shown in Fig.~\ref{sfh_Lagos}. However, to serve as a fairer comparison to the case where MSI 
is applied to observational data, we instead choose a simple parametric form for $\psi(M_\star,t)$ 
given by

\begin{align}
\frac{\langle\psi(M_\star,t)\rangle}{\mathrm{M_\odot Gyr^{-1}}} = 10^{11} \, \left(\frac{c(t)}{\mathrm{Gyr^{-1}}}\right) \left(\frac{M_\star}{10^{11} \, \mathrm{M_\odot}}\right)^{1+\beta_{\mathrm{sf}}},
\label{sfr_seq_galform}
\end{align}

\noindent where $\beta_{\mathrm{sf}}$ is the power-law slope of the star forming sequence which is assumed 
to be constant with time. $c(t)$ specifies the evolution in the normalisation 
of the star forming sequence. We find that a reasonable parametrisation for the normalisation 
is given by fitting a power law of the form

\begin{equation}
c(t) = 0.95 \, (1+z)^{1.23} \, \mathrm{Gyr^{-1}}.
\label{sfr_seq_fit_eqn}
\end{equation}

We note that this simple parametrisation of $\langle\psi(M_\star,t)\rangle$ is clearly an oversimplification 
given that the predicted power-law slope, $\beta_{\mathrm{sf}}$, of the sequence shown in Fig.~\ref{ssfr_m_evo_Lagos12} 
and Fig.~\ref{ssfr_m_evo_Lagos_obs_comp} steepens with redshift. However, the slope inferred
from the observational data described in Appendix~\ref{MSI_Obs_Section} is not sufficiently 
well constrained with regard to showing a convincing evolution with redshift. Therefore, for the 
purposes of making a fair assessment of the MSI technique when applied to observational data, we 
choose to keep $\beta_{\mathrm{sf}}$ as constant in time.

The result of applying MSI to our fiducial \galform model can be seen by comparing the blue 
(intrinsic) and green (inferred from MSI) lines in Fig.~\ref{sfh_Lagos}. The agreement is not 
perfect. However, it can be seen that MSI broadly reproduces the flat shape of the stellar mass 
assembly histories predicted by our fiducial model. The worst agreement is seen for the 
$\log(M_\star(t_0) / \mathrm{M_\odot}) = 8.5$ bin, where MSI predicts that the mass assembly 
rate should steadily rise from early times up to $z=0$. In addition, our application of MSI 
slightly underpredicts the mass assembly rates of the most massive galaxies, such that the predicted
stellar mass assembly histories do not drop correctly at early times.

We now consider several potential shortcomings of the MSI technique that could all contribute to 
this disagreement. Firstly, MSI assumes that star forming galaxies at a given point in time have 
always been on the star forming sequence prior to that time. We showed in Fig.~\ref{ssfr_m_evo_Lagos12} 
that there is a tight star forming sequence predicted 
by our fiducial model. However, in principle it is possible that galaxies could be quenched (for 
example by a major merger triggered starburst event using up all the cold gas) before accreting enough fresh gas onto a disk 
to rejoin the star forming sequence. The impact from such a scenario can be tested in a 
straightforward manner by considering the dispersion in the distribution of mass assembly rates around the mean, as shown
in Fig.~\ref{sfh_Lagos}. We find that the typical dispersion is roughly compatible with the 
dispersion at a given stellar mass of the star forming sequence shown in 
Fig.~\ref{ssfr_m_evo_Lagos12}. This supports, in a statistical sense, the assumption 
folded into the MSI technique that galaxies which are star forming at $z=0$ do not drop below the 
sequence at some earlier stage in their evolution, at least for a significant period of time. 

A second potential shortcoming of the MSI technique is that it ignores the hierarchical assembly 
of stellar mass through galaxy merging events. It is possible that a significant fraction of the stellar 
mass of a star forming galaxy at $z=0$ was formed in multiple progenitors, in which case the MSI method 
breaks down unless the sum of these progenitors also conspires to reside on the star forming sequence. We 
check for the contribution from merging by comparing the mean stellar mass assembly histories of the main 
stellar progenitors (solid blue lines) to the sum of all stellar progenitors (black lines) in 
Fig.~\ref{sfh_Lagos}. Over all of the stellar mass bins considered, it can be seen from Fig.~\ref{sfh_Lagos} 
that the stellar mass assembly histories of galaxies that are central and still star forming at $z=0$ are dominated 
by the main stellar progenitor, providing support for the validity of the MSI technique. 
This result is perhaps unsurprising, in that in order for the stellar mass of a secondary progenitor
to become significant relative to the stellar mass of the main progenitor, the system must undergo a major 
merging event which would ultimately quench star formation in the resulting galaxy as gas is used 
up in a starburst event.

\begin{figure}
\includegraphics[width=20pc]{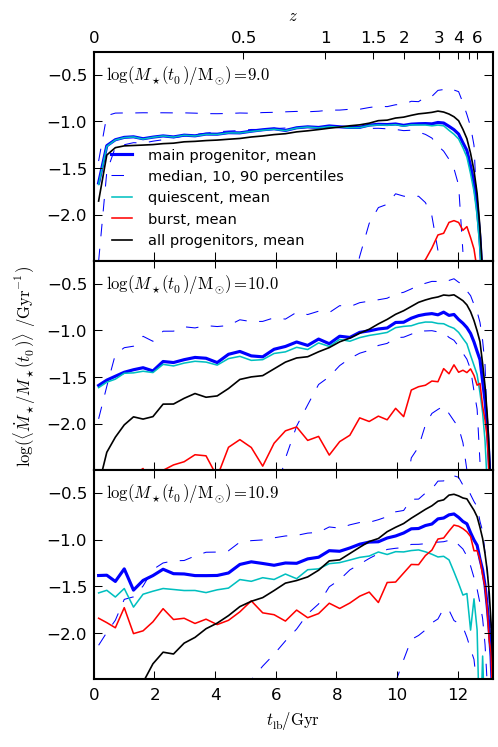}
\caption{The average stellar mass assembly histories of central galaxies that are passive at $z=0$ from our fiducial \galform model, plotted as a function of lookback time.
Model galaxies are binned by their stellar mass at $z=0$, with each panel corresponding to a different mass bin.
The median $z=0$ stellar mass in each bin is labelled in each panel.
The meaning of the lines is the same as for \protect Fig.~\ref{sfh_Lagos}.}
\label{sfh_passive}
\end{figure}

To emphasise the difference between the star forming galaxy population we consider here and passive 
galaxies, we show in Fig.~\ref{sfh_passive} the average stellar mass assembly histories of model galaxies that are central and passive 
at $z=0$. In contrast to Fig.~\ref{sfh_Lagos}, the stellar mass assembly 
histories of the main progenitors of passive centrals (blue lines) are significantly different from the stellar 
mass assembly histories obtained from summing over all progenitors (black lines). This difference is largest for 
the most massive galaxies where a significant amount of stellar mass is assembled in secondary 
progenitors at early times which merge onto the main progenitor galaxy later. The blue lines also 
include the rate of accretion of stellar mass from secondary progenitors and can therefore exceed the black 
lines in this case.

Returning to the star forming galaxy sample, Fig.~\ref{sfh_Lagos} also shows that quiescent star 
formation (cyan lines) mostly dominates the stellar mass assembly rates of galaxies which are still 
star forming at $z=0$, as compared to star formation in bursts (red lines). The only exception 
to this is for the progenitors of massive star forming galaxies at $z=0$, where bursts briefly 
dominate the stellar mass assembly process at high redshift. Integrated over the lifetime of these 
galaxies however, the burst star formation mode is still entirely subdominant. This is important 
for the MSI technique because bursts can perturb galaxies above the star forming sequence. 
However, as has also been shown by \cite{Lamastra13}, we find 
that actively bursting galaxies in hierarchical galaxy formation models can also reside on (or in 
same cases below) the star forming sequence. 
As an aside, the result that star formation in the galaxies considered in Fig.~\ref{sfh_Lagos} 
is dominated by quiescent star formation in the main stellar progenitor lends support to the 
methodology employed by galaxy formation models which ignore galaxy merging and disk instabilities,
provided that these models are used only to predict the statistical properties of actively 
star forming central galaxies \cite[e.g.][]{Dutton10}.

Finally, it should be noted that the MSI technique which we employ for this study includes the 
assumption that the star forming sequence can be described by a single, unbroken power law over 
all relevant scales in stellar mass. Fig.~\ref{ssfr_m_evo_Lagos_obs_comp} shows that this is only 
approximately true for the star forming sequence in our fiducial \galform model. If the true
star forming sequence cannot be adequately described by a single power law then the resulting stellar 
mass assembly histories inferred using the MSI technique will be in error. As we start the 
integration process at $z=0$, this error would become more severe at early times. 
In addition, as noted earlier, Fig.~\ref{ssfr_m_evo_Lagos_obs_comp} shows that the power-law slope of 
the star forming sequence, $\beta_{\mathrm{sf}}$, evolves with redshift in our fiducial model.
Comparison of the true (solid blue) and inferred (green) average stellar mass assembly histories in 
Fig.~\ref{sfh_Lagos} shows that MSI does not perfectly agree with the direct model prediction, and
that the disagreement becomes worse at early times. Given that the other potential sources of error 
which we have considered until now appear to be insignificant, we attribute the disagreement between
MSI and the direct model output seen in Fig.~\ref{sfh_Lagos} to the simple power-law parametrisation of 
$\langle\psi(M_\star,t)\rangle$ which we use to perform MSI.
Given these problems, any comparison between MSI and direct model predictions should only be 
interpreted taking into account that the MSI technique likely fails to precisely constrain the 
shape of the stellar mass assembly histories of galaxies, particularly at early times. Nonetheless, 
the qualitative trend of almost flat stellar mass assembly histories seen in Fig.~\ref{sfh_Lagos} is 
broadly reproduced by the MSI technique for all but the least massive galaxies. We can therefore 
proceed to perform a qualitative comparison between the shapes of the stellar mass assembly histories 
predicted by our model and those inferred from observational data using MSI.

\subsection{Applying main sequence integration to observational data}
\label{MSI_Obs_Section}

\begin{table*}
\begin{center}
\begin{tabular}{|cccccc|}
\hline
Source & Redshift& Selection& SF cut& Tracer& Symbol \\
\hline

\cite{Daddi07}&       1.4-2.5& BzK&          sBzK&                        UV (corrected)&           \textcolor{magenta}{$\times$}\\
\cite{Elbaz07}&       0.8-1.2& z&            blue colour&                 $24 \mathrm{\mu m}$+UV&   \textbf{$\times$}\\
\cite{Salim07}&       0.05-0.2& r&           BPT diagram&                 SED fitting&              {$\times$}\\
\cite{Santini09}&     0.3-2.5& $K_{\mathrm{s}}$& SFR-$M_\star$ distribution&      $24 \mathrm{\mu m}$+UV&   \textbf{$\blacklozenge$}\\
\cite{Labbe10}&       7&       LBG&          blue colour&                 UV (corrected)&           \textcolor{green}{$\blacksquare$}\\
\cite{Oliver10}&      0-2&     Optical&      template fitting&            $70/160 \mathrm{\mu m}$&  \textcolor{blue}{$\blacklozenge$}\\
\cite{Peng10}&        0-1&     Optical&      blue colour&                 SED fitting&              \textcolor{blue}{$\times$}\\
\cite{Rodighiero10}&  0-2.5&   $4.5 \mathrm{\mu m}$& blue colour/$24 \mathrm{\mu m}$ detection& FIR& $\bigstar$\\ 
\cite{Elbaz11}&       0-3&     $24 \mathrm{\mu m}$ & $24 \mathrm{\mu m}$ detection& FIR&            $\blacktriangleright$\\
\cite{Karim11}&       0.2-3&   $3.6 \mathrm{\mu m}$& blue colour&         Radio&                    \textbf{$\bullet$}, \textcolor{blue}{$\bullet$}\\
\cite{Bouwens12}&     4&       LBG &         blue colour&                 UV (corrected)&           \textbf{$+$}\\
\cite{Lin12}&         1.8-2.2&       BzK&          sBzK&                        UV (corrected)&     $\blacktriangle$\\
\cite{Reddy12}&       1.4-3.7& LBG&          blue colour&                 $24 \mathrm{\mu m}$+UV&   \textcolor{blue}{$+$}\\
\cite{Sawicki12}&     2.3&     UV&           blue colour&                 UV (corrected)&           $\blacksquare$\\
\cite{Whitaker12}&    0-2.5&   K&            (U-V/V-J) cut&               $24 \mathrm{\mu m}$+UV&   $+$\\
\cite{Koyama13}&      0.4,0.8,2.2& $H_\alpha$& $H_\alpha$&                $H_\alpha$ (corrected)&   \textcolor{green}{$\blacklozenge$}\\
\cite{Wang13}&        0.2-2&   K&            SFR-$M_\star$ distribution&          SED fitting / FIR&        \textcolor{blue}{$\blacksquare$}\\

\end{tabular}
\caption{List of the sources of power-law fits to the observed star forming sequence extracted from the literature.
We list the source, redshift range or median redshift, galaxy selection technique, the subsequent star forming galaxy selection technique, and the tracer used to estimate the instantaneous star formation rate.
The symbols used for each source in Fig.~\ref{beta_tlb} and Fig.~\ref{logc_tlb} are also shown.
For LBG selected samples, it should be noted that the initial galaxy selection technique is strongly biased towards blue star forming galaxies, so typically no additional cut to separate star forming galaxies is performed.
For \protect\cite{Karim11}, we use both the star forming galaxy sample presented in their Table 3 as well as the active population which is shown in their Figure 13 (which uses a bluer colour cut).
The code and observational data used for this compilation are available at \protect\url{http://www.astro.dur.ac.uk/~d72fqv/MS_fits/}.}
\label{ms_fit_table}
\end{center}
\end{table*}

To infer the average stellar mass assembly histories of galaxies from observations using the MSI technique, it is necessary to 
specify $\langle\psi(M_\star,t)\rangle$ for all possible values of $M_\star$ and $t$. Rather than attempt to 
interpolate directly between the observational data on the average specific star formation rates presented in 
Section~\ref{ssfr_model_obs_comp_section}, we instead choose to first compile a list of power-law fits 
to the star forming sequence for different redshifts from the literature. Basic information on this 
compilation is presented in Table~\ref{ms_fit_table}. Using Equation~\ref{sfr_seq_galform}, we parametrise 
these power-law fits with the slope, $\beta_{\mathrm{sf}}$ and the normalisation, $c(t)$.
We convert the fits taken from studies that assume a Salpeter IMF to a Chabrier IMF by applying a 
correction of $-0.24 \, \mathrm{dex}$ to both $\psi$ and $M_\star$ \citep{Ilbert10,Mitchell13}. This typically makes only a very
small difference to the resulting power-law fits.

\begin{figure}
\includegraphics[width=20pc]{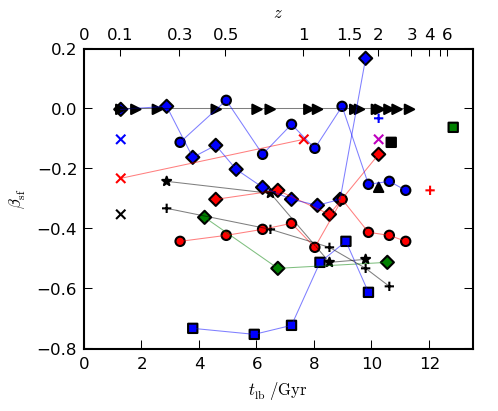}
\caption{The slope of power-law fits to the observed star forming sequence from the literature, plotted as a function of lookback time.
Each symbol corresponds to data from a different source.
The list of sources for the compilation is presented in \protect Table~\ref{ms_fit_table}, which also references which source matches a given symbol.}
\label{beta_tlb}
\end{figure}

We show our observational compilation of $\beta_{\mathrm{sf}}$ as a function of lookback time in Fig.~\ref{beta_tlb}. This shows 
that currently there is not a strong consensus on the slope of the star forming sequence in the 
literature. Given the wide range of selection techniques that are used to separate 
star forming galaxies, we expect the variation in $\beta_{\mathrm{sf}}$ seen in Fig.~\ref{beta_tlb} 
to be driven primarily by selection effects. For example, \cite{Karim11} explore this issue in an appendix and 
show that increasingly blue rest-frame $(NUV - r)$ colour cuts result in increased values of 
$\beta_{\mathrm{sf}}$. Another issue is whether the star forming sequence can really be described by a single 
unbroken power law in $M_\star$ \cite[see][]{Huang12}. For example, if the slope of the sequence changes at the
high mass end then the range in stellar mass probed by each individual study will have an effect on the 
inferred slopes. Inspection of Fig.~\ref{ssfr_m_evo_Lagos_obs_comp} shows evidence that this does indeed
occur in our fiducial model.

\begin{figure}
\includegraphics[width=20pc]{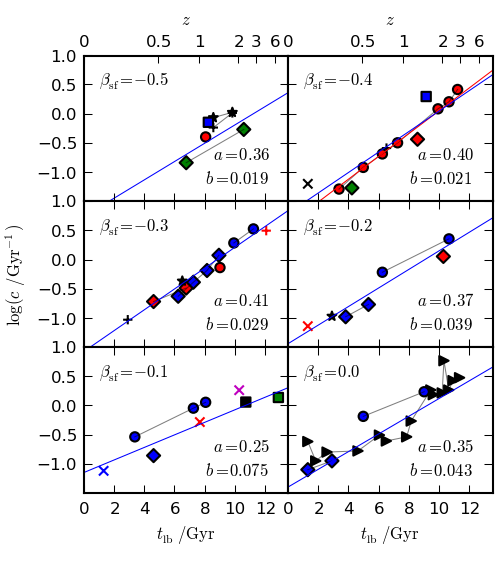}
\caption{Normalisation of power-law fits to the star forming sequence from the literature, plotted as a function of lookback time.
The list of sources for the compilation is presented in \protect Table~\ref{ms_fit_table}, which also references which source matches a given symbol.
Each panel shows the normalisation for different bins of the fitted power-law slope to the star forming sequence, $\beta_{\mathrm{sf}}$, as labelled.
For each panel, the evolution of the normalisation is fitted by $(c/\mathrm{Gyr^{-1}}) = b \, \exp(a \, t_{\mathrm{lb}} / \mathrm{Gyr})$ and the best fitting $a$ and $b$ are labelled.
These fits are shown as blue lines and use all of the observational data, including data from \protect\cite{Karim11}.
In addition, we also perform an independent fit (red line) to just the \protect\cite{Karim11} star forming sample (red circles) in isolation for the $\beta_{\mathrm{sf}} = -0.4$ bin.}
\label{logc_tlb}
\end{figure}

Given this uncertainty in the true slope of the star forming sequence, we first make the simplest possible
assumption, which is that the slope remains constant with lookback time. We then choose to estimate 
$\langle\psi(M_\star,t)\rangle$ by first binning the power-law fits from Table~\ref{ms_fit_table} in $\beta_{\mathrm{sf}}$, before 
performing a fit to $c(t)$ for each bin as a function of lookback time. For two of the 
studies included in our compilation \citep{Lin12, Reddy12}, a best fitting slope to the star forming sequence
is provided but the corresponding normalisation is not available, and so they do not appear in 
Fig.~\ref{logc_tlb} or feature in the following fits. The resulting data and fits to the evolution in the
normalisation are shown in Fig.~\ref{logc_tlb}. 

Unlike the slope, there is actually a reasonable consensus in the literature 
on the evolution in $c(t)$ for a given $\beta_{\mathrm{sf}}$ bin. We find that the evolution in the 
normalisation seen in Fig.~\ref{logc_tlb} is best fit as an exponential function of lookback time rather than as a
power law in $(1+z)$. We therefore parametrise the evolution in the normalisation using 

\begin{equation}
\frac{c(t)}{\mathrm{Gyr^{-1}}} = b \, \exp{\left(a \, \frac{t_{\mathrm{lb}}}{\mathrm{Gyr}}\right)}.
\label{logc_evo_data}
\end{equation}

\noindent To account for the oversampling in the number of points at $z \approx 2$ in some of the $\beta_{\mathrm{sf}}$ bins, we 
weight all the points shown in Fig.~\ref{logc_tlb} to give equal weight to each bin in $\Delta t_{\mathrm{lb}} = 1 \, \mathrm{Gyr}$
within each panel. In order to facilitate a qualitative comparison with the method used by \cite{Leitner12} to estimate the star
formation histories of star forming galaxies, we also perform an independent fit to the star forming 
galaxy sample from \cite{Karim11}, fixing $\beta_{\mathrm{sf}} = -0.4$. This fit to the evolution in the normalisation, 
$c(t)$, is shown by the red line in Fig.~\ref{logc_tlb}.

\begin{figure*}
\begin{center}
\includegraphics[width=40pc]{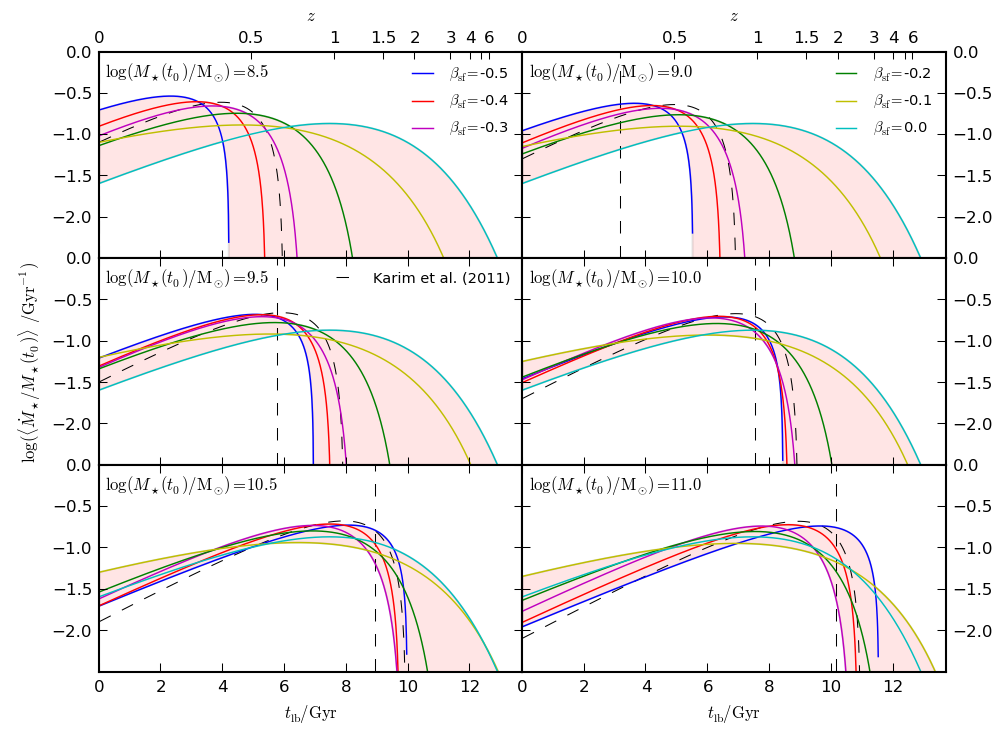}
\caption{The stellar mass assembly histories of star forming galaxies inferred by applying MSI to observational data, plotted as a function of lookback time.
Each panel corresponds to a different $z=0$ stellar mass, as labelled.
Coloured solid lines show the mass assembly histories inferred by applying MSI to the observational compilation presented in \protect Table~\ref{ms_fit_table}.
Each solid line in a given panel corresponds to a different bin in the power-law slope of the star forming sequence, $\beta_{\mathrm{sf}}$, as taken from the compilation.
Dashed black curves show the mass assembly histories inferred by applying MSI to the star forming galaxy sample from \protect\cite{Karim11}.
The dashed vertical lines show the lookback time beyond which the MSI technique, applied to the \protect\cite{Karim11} sample, extrapolates below the stellar mass completeness limits of \protect\cite{Karim11}.
For the $\log(M_\star(t_0) \, / \mathrm{M_\odot}) = 8.5$ panel, the entire stellar mass assembly history inferred from \protect\cite{Karim11} involves an extrapolation below this mass completeness limit.}
\label{sfh_obs}
\end{center}
\end{figure*}

Once $\langle\psi(M_\star,t)\rangle$ has been parametrised, we can apply MSI to infer the average stellar mass assembly 
histories of $z=0$ star forming galaxies for different values of the stellar mass at $z=0$, $M_\star (t_0)$. 
The results of this exercise are shown as coloured lines in Fig.~\ref{sfh_obs}, with each line corresponding
to a different bin in $\beta_{\mathrm{sf}}$. To compare with the approach used by \cite{Leitner12}, we also apply MSI to 
only the star forming sample presented in \cite{Karim11}. The results of doing this are shown by the dashed 
black curves in Fig.~\ref{sfh_obs}.

It is immediately apparent that the uncertainty on the slope of the star forming sequence reported in the 
literature translates to a considerable uncertainty on the stellar mass assembly histories inferred 
from the data. The uncertainty is largest for low mass galaxies where, in particular, the formation time at 
which a given galaxy forms a given fraction of its stars is very poorly constrained. This partly reflects the fact 
that an increasingly large extrapolation in $\langle\psi(M_\star,t)\rangle$ has to be made for smaller galaxies as the stellar 
mass of their progenitors typically drops below the completeness limit of the observational surveys used to 
obtain $\langle\psi(M_\star,t)\rangle$.

Despite the considerable uncertainties, qualitatively the data seems to favour a 
scenario where galaxies that are still star forming at $z=0$ undergo a peak phase of star formation activity 
at $z \approx 1$, followed by a drop towards late times. The actual position of the peak and the rate of late 
time decline are somewhat poorly constrained. Furthermore, for $\beta_{\mathrm{sf}} < -0.2$, the position of this peak clearly 
depends on $M_\star (t_0)$, such that a downsizing trend is apparent. Massive star forming galaxies are 
inferred to form a greater fraction of their stellar mass at early times compared to lower mass galaxies in this 
case. This is the conclusion presented in \cite{Leitner12} who use only data from \cite{Oliver10} and 
the star forming sample from \cite{Karim11} as inputs to their application of MSI. Such a downsizing trend, 
provided that star forming galaxies are successfully separated from passive galaxies, should be completely 
independent of any physical processes that cause permanent quenching of star formation. On the other hand, if $\beta_{\mathrm{sf}}$ is 
larger, then the shapes of the stellar mass assembly histories of galaxies that are star forming at $z=0$ are almost 
completely independent of their final stellar mass.

The stellar mass assembly histories inferred from applying MSI to only the star forming sample presented in 
\cite{Karim11} (dashed black lines) are mostly consistent with the curves obtained by fitting to data taken 
from the entire observational compilation. This implies that the approach used by \cite{Leitner12}, which relies
primarily on the \cite{Karim11} data, should yield results that are consistent with ours. On the other hand, it can be seen from the red line shown in 
Fig.~\ref{logc_tlb} that extrapolating the \cite{Karim11} results down to $z=0$ favour a steeper late time 
drop in the normalisation of the star forming sequence than is implied by SDSS data \citep{Salim07}. This is reflected in 
the steeper drop in the stellar mass assembly histories inferred from applying MSI to only the \cite{Karim11} 
data in Fig.~\ref{sfh_obs}. This emphasises the need to consider results from as much of the literature as 
possible in order to try to account for the considerable uncertainties on the slope and normalisation of 
the star forming sequence.

\section{Invariance In The Shape Of Predicted Stellar Mass Assembly Histories}
\label{Invariance_Section}

In Section~\ref{sfh_model_obs_comp}, we demonstrated that, qualitatively, the stellar mass assembly 
histories predicted by our fiducial \galform model do not agree closely with the trends we infer from 
observational data. In this appendix, we address the fact that our fiducial model is only one specific realisation of \galform with
regard to the various model parameters that can be changed. These parameters are constrained by matching global diagnostics
of the galaxy population. We now proceed to demonstrate that the disagreement between \galform and 
the observational data holds for a wide range of choices of these model parameters.
This result stems from our finding that the shapes of the average stellar mass assembly histories of central 
star forming galaxies in \galform are almost entirely invariant when changing model parameters relating to 
star formation, feedback and gas reincorporation.

\begin{figure*}
\begin{center}
\includegraphics[width=40pc]{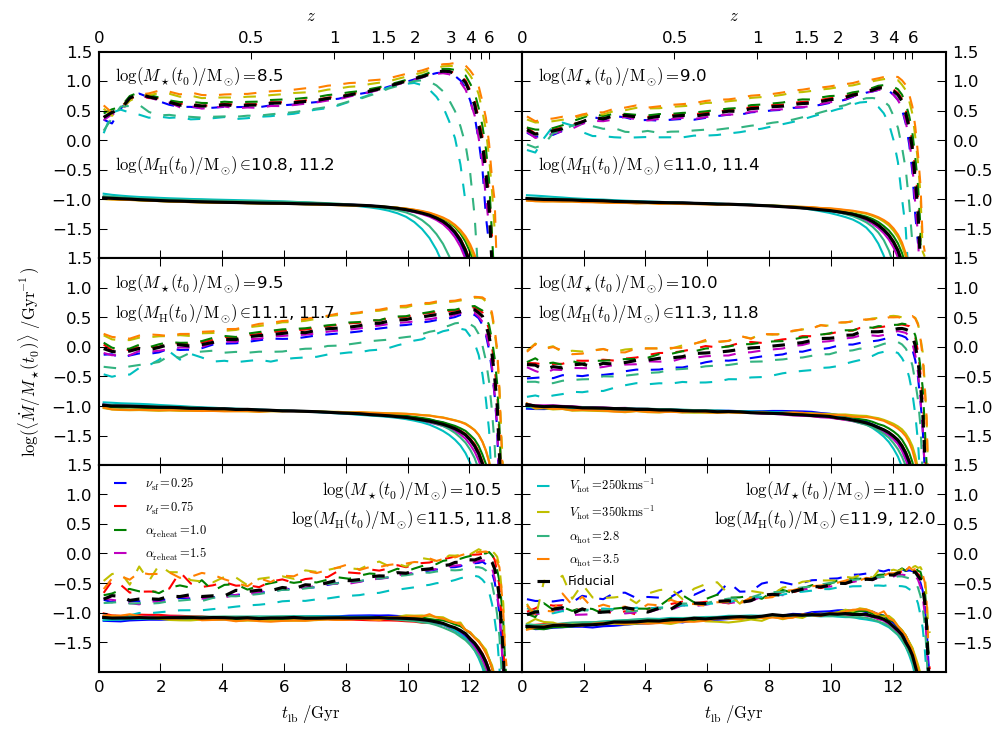}
\caption{The mean mass assembly histories of model central galaxies that are star forming at $z=0$, plotted as a function of lookback time.
Solid lines show predictions for the mean stellar mass assembly histories of the main stellar progenitors of galaxies.
Black lines correspond to our fiducial \galform model.
Other colours correspond to variations of our fiducial model, with a single model parameter changed to the labelled value.
Definitions of these model parameters can be found in Section~\ref{galform_feedback}.
Dashed lines show the corresponding dark matter halo mass assembly histories of the progenitor haloes that host the main stellar progenitors of central star forming galaxies at $z=0$.
The halo mass assembly curves are rescaled by $\Omega_{\mathrm{b}}/\Omega_{\mathrm{M}}$ to show the baryonic accretion rate onto these haloes.
Model galaxies are binned according to their $z=0$ stellar mass with each panel corresponding to a different mass bin.
The median $z=0$ stellar mass in each bin is labelled in each panel.
The range (across all of the \galform models shown) in the corresponding median $z=0$ dark matter halo mass of each stellar mass bin is also labelled.}
\label{sfh_invariance}
\end{center}
\end{figure*}

We demonstrate this behaviour in Fig.~\ref{sfh_invariance}, which shows the average stellar mass and 
halo mass assembly histories of model galaxies which are central and star forming at $z=0$. We show 
the output of a variety of variants of our fiducial model. These variants are chosen as examples to display the 
range of mass assembly histories which arise as a result of changing various model parameters in 
\galform which are relevant to star forming galaxies. Included are parameters that control the global
efficiency of star formation and SNe feedback, the gas reincorporation timescale and the dependence of
the mass loading factor, $\beta$, on the circular velocity of galaxies.
Note that it is possible, in principle, 
that changing these model parameters would affect the position of the star forming sequence in \galform, 
invalidating our separation between star forming and passive galaxies. We have verified that this is, in fact, not 
the case and find that the star forming galaxy cuts shown as blue lines in Fig.~\ref{ssfr_m_evo_Lagos12} 
continue to be effective at isolating the star forming sequence for all the models and redshifts considered 
here.

To first order, the stellar mass assembly histories for all the models shown in Fig.~\ref{sfh_invariance} 
are almost identical for a given stellar mass, $M_\star(t_0)$, with significant variations only occurring at early times. In contrast, the 
normalisation of the halo mass assembly histories shifts significantly between different choices of model
parameters. Therefore, while these model parameters in \galform are capable of changing the $z=0$ stellar 
mass function by changing the relationship between stellar mass and halo mass, they do not significantly
affect the stellar mass assembly process of central galaxies of a given stellar mass at $z=0$.

This result can be understood by first reviewing the way that the stellar mass assembly process takes place in 
\galform. As discussed in, for example, \cite{Fakhouri10}, the specific halo mass assembly rate and 
consequently the shape of the corresponding dark matter halo mass assembly histories are nearly independent 
of the final halo mass (see their Equation 2). This can be seen directly in Fig.~\ref{sfh_invariance}. 
Secondly, as shown in Section~\ref{sfh_model_obs_comp}, stellar mass assembly broadly tracks halo mass 
assembly in our fiducial \galform model. As described in Section~\ref{Explain_SFH_Section}, this coevolution
arises because the mass loading and reincorporation efficiencies do not evolve significantly over the majority of the
lifetimes of typical star forming galaxies. For the parametrisations currently used to model these physical
processes in \galform, changing the relevant model parameters merely changes their efficiency in a global sense. 
Therefore, in order to change the shape of the stellar mass assembly histories, an alternative parametrisation
of one (or both) of these processes would be required. Such a modification would need to result in significantly
stronger evolution in the mass loading factor, $\beta_{\mathrm{ml}}$, or the reincorporation timescale, 
$t_{\mathrm{ret}}$, than is seen for our fiducial model in Fig.~\ref{timescales}.

\section{Virial Mass Scaling Model}
\label{H12_section}

In Section~\ref{tret_mod_section}, we demonstrate that by modifying the reincorporation timescale in the model to the form 
given by Eqn~\ref{reincorporation_modified}, it is possible to reconcile the predicted and inferred stellar mass assembly histories
of galaxies that are still star forming at $z=0$. In Fig.~\ref{tret_modified} we show that this ad hoc model for the 
reincorporation timescale is quite different from the modification introduced by \cite{Henriques13} in order
to reproduce the observed evolution of the stellar mass and luminosity functions. We now consider implementing
into our model their suggestion that the reincorporation timescale should only scale with the halo virial mass as

\begin{equation}
t_{\mathrm{ret}} = \gamma \, \frac{10^{10} \mathrm{M_\odot}}{M_{\mathrm{H}}} .
\label{tret_bruno}
\end{equation}

We start by requiring that this virial mass scaling model should provide an adequate match to the $z=0$ stellar
mass function. Starting from our fiducial model, we find that it is possible to do this simply by implementing Eqn~\ref{tret_bruno}
with $\gamma$ changed from $18 \, \mathrm{Gyr}$, as in \cite{Henriques13}, to $5.6 \, \mathrm{Gyr}$.

Comparisons with the other models considered in this paper for the evolution in the specific star formation rates
and stellar mass assembly histories of star forming galaxies are shown in Fig.~\ref{ssfr_zm_Lagos_obs_comp} and Fig.~\ref{sfh_modified}
respectively. Fig.~\ref{sfh_modified} shows that the virial mass scaling model predicts rapidly rising stellar mass assembly
histories for all but the most massive star forming galaxies at $z=0$. For galaxies with $\log(M_\star / \mathrm{M_\odot}) = 10$
at $z=0$, this results in dramatic disagreement with the stellar mass assembly histories inferred from observations with too
much star formation at late times compared to early times. For low mass galaxies, the observational constraints are very weak,
such that any model could be compatible. For the most massive systems, we note that similar to the modified reincorporation
model shown in red, the cooling timescale in the virial mass scaling model becomes long compared the reincorporation 
timescale for galaxies residing within massive haloes. As discussed in Section~\ref{tret_mod_section}, modifying the reincorporation
timescale in this regime will have a much smaller impact as the gas cycle becomes limited by cooling. Fig.~\ref{ssfr_zm_Lagos_obs_comp} shows
that the virial mass scaling model predicts specific star formation rate evolution which is too slow at a fixed stellar mass
relative to the rapid evolution inferred from observations. Again, the specific star formation rates of massive star forming 
galaxies are relatively unaffected by modifications to the reincorporation timescale.

\begin{figure*}
\begin{center}
\includegraphics[width=40pc]{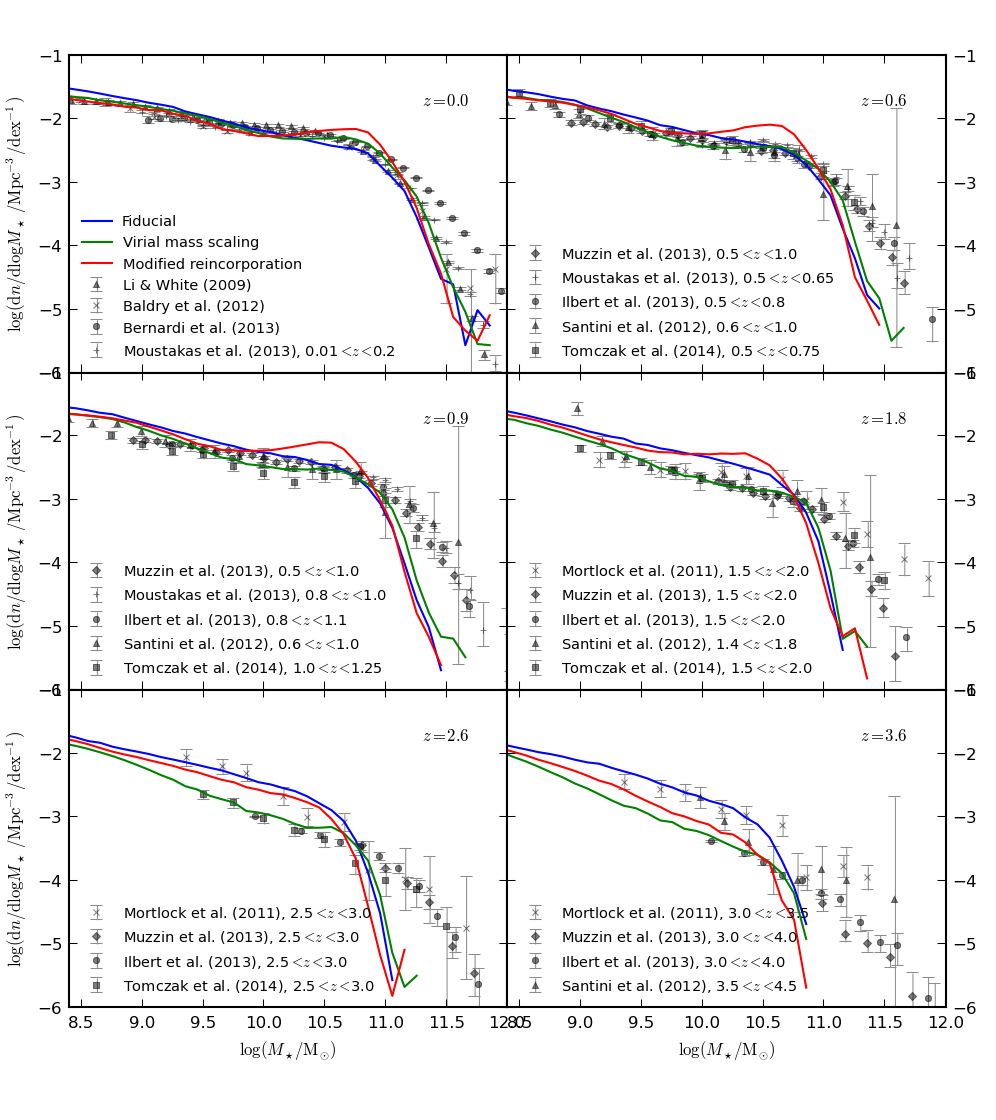}
\caption{Stellar mass functions predicted by the different \galform models for a selection of redshifts, as labelled in each panel.
Blue lines show predictions from our fiducial model.
Red lines correspond to the modified reincorporation model introduced in \protect Section~\ref{tret_mod_section}.
Green lines correspond to a model using the virial mass scaling for the reincorporation timescale introduced by \protect \cite{Henriques13}.
The grey points and error bars show observational estimates of the stellar mass function from \protect \cite{Li09}, \protect \cite{Baldry12}, \protect \cite{Ilbert10}, \protect \cite{Mortlock11}, \protect \cite{Santini12},
\protect \cite{Bernardi13}, \protect \cite{Ilbert13}, \protect \cite{Moustakas13}, \protect \cite{Muzzin13b} and \protect \cite{Tomczak14}. 
Where necessary we convert these observational results from a Salpeter to a Chabrier IMF using a $-0.24\,\mathrm{dex}$ correction \protect \citep{Ilbert10,Mitchell13}.}
\label{mf_evo}
\end{center}
\end{figure*}

While our modified reincorporation model is more successful than the virial mass scaling model in reproducing the trends
inferred from the data in Fig.~\ref{ssfr_zm_Lagos_obs_comp} and Fig.~\ref{sfh_modified}, it is important to also consider the predicted evolution
of the stellar mass function, given that this is the main constraint used by \cite{Henriques13}. Stellar mass function
predictions for the different models considered in this paper are compared to a compilation of observational data in Fig.~\ref{mf_evo}.
To first order, the differences between the models can be summarised simply by noting that relative to our fiducial model,
the modified reincorporation model suppresses both early and late star formation. The virial mass scaling model also suppresses early
star formation but predicts much stronger star formation at late times for all but the most massive galaxies. This results
in much stronger evolution in number density short of the break in the mass function below $z=2$, in good agreement with the most recent observational
data. In contrast, our fiducial model predict an overabundance of low mass galaxies relative to the observations beyond the local
Universe.

Compared to the observations, all three models shown here all strongly underpredict the abundance 
of galaxies above the break of the stellar mass function beyond the local Universe.
The extreme end of the mass function must always be interpreted with care because of the potential for Eddington bias to artificially boost
the population residing in the massive tail of the galaxy stellar mass distribution. On the other hand, the disagreement at the
massive end could be related to the problem that all three models suffer from in failing to reproduce the rapid evolution in the
specific star formation rates of massive star forming galaxies with $\log(M_\star / \mathrm{M_\odot}) = 11$  seen in Fig.~\ref{ssfr_zm_Lagos_obs_comp}.

At this stage, it is tempting to speculate whether it is possible to produce a model that can be consistent with
the observed stellar mass function and star formation evolution simultaneously. One possibility that we did not fully explore in this
study would be to search for a modified model which introduces a strong downsizing trend into the purely star forming
population, such that low mass galaxies form very late, and yielding a star forming sequence with a slope of $\beta_{\mathrm{SF}} \approx -0.4$,
compatible with current observations of the star forming sequence. In this case, it might be possible to reproduce the observed evolution
in the low mass end of the stellar mass function while also predicting peaked star formation histories for star forming galaxies with
$\log(M_\star / \mathrm{M_\odot}) = 10$ at $z=0$. We defer any further exploration of this issue to a future paper.

Another consideration is that currently it is assumed in our model that all gas which is removed from galaxies by SNe feedback 
is added to a reservoir instead of being added straight back into the hot halo gas profile. In the case where the
reincorporation timescale is of order the halo dynamical time, this is equivalent to assuming that the ejected gas moves in a ballistic fashion
and escapes the halo virial radius before returning. In contrast, the models presented by
\cite{Guo11} and \cite{Henriques13} assume that only a fraction of this gas is actually ejected from the halo. The remaining fraction is added
back into the halo gas profile and is therefore able to inflow back onto the disk very rapidly when cooling timescales are short. In these
models, the fraction of gas that is able to escape the halo is parametrised as a function of the halo circular velocity, $V_{\mathrm{vir}}$,
such that a larger fraction of gas is ejected from the haloes of low mass systems. Depending on the set of chosen model parameters, this
treatment of the gas affected by SNe feedback could change the $z=0$ stellar mass range over which a modification to the reincorporation timescale
would be effective in changing the stellar mass assembly histories of star forming galaxies. Again, we defer any further exploration in this area to future work.

\label{lastpage}
\end{document}